\documentclass[a4paper,12pt,preprint,nofootinbib]{revtex4}
\renewcommand{\vec}[1]{\mathbf{#1}}
\usepackage{cancel}
\usepackage{amsmath,mathrsfs,amscd,graphicx}
\usepackage{mciteplus}
\usepackage{multirow}
\usepackage{color}
\usepackage{cancel}
\usepackage{rotating}
\usepackage{slashed}
\usepackage{subfig}
\usepackage{textcomp}
\usepackage[final]{pdfpages}
\usepackage{array}
\usepackage{float}
\usepackage{url}
\usepackage{textcomp}
\usepackage{amsfonts, latexsym, epsfig}
\usepackage{bm}
\usepackage{times}
\usepackage{epsfig}
\usepackage{amssymb}
\usepackage{tikz}
\usepackage{slashed}
\usepackage{hyperref}
\usepackage{verbatim}
\usepackage[normalem]{ulem}
\usepackage[utf8]{inputenc}
\usepackage{diagbox}
\usepackage{titlesec}
\def\beq{\begin{equation}}
\def\eeq{\end{equation}}
\def\be{\begin{equation}}
\def\ee{\end{equation}}
\def\bea{\begin{eqnarray}}
\def\eea{\end{eqnarray}}

\definecolor{dpmagenta}{rgb}{0.8, 0.0, 0.8}

\begin{document}
\title{Photo-production of the Higgs Boson at the LHeC}
\author{Ruibo Li~}
\email{bobli@zju.edu.cn}
\author{Xiang Lv~}
\email{lvxiang@zju.edu.cn}
\author{Bo-Wen Wang~}
\email{0617626@zju.edu.cn}
\author{Kai Wang~}
\email{wangkai1@zju.edu.cn}
\author{Tao Xu}
\email{taoxu.4@gmail.com}
\affiliation{Zhejiang Institute of Modern Physics and Department of Physics, Zhejiang University, Hangzhou, Zhejiang 310027, China 
}
\begin{abstract}
	As one category of vector boson fusion, photo-production is one important  production mechanism at $e$-$p$ colliders. A  future $e$-$p$ collider -- Large Hadron-electron Collider (LHeC) has been discussed as a ``Higgs factory'' candidate where the Higgs boson produced via weak boson fusion (WBF) at the LHeC plays an important role in precision measurement of Yukawa couplings. On the other hand, a measurement of photo-production of the Higgs boson, if possible, might be complementary to the measurement of Higgs to di-photon partial decay width $\Gamma({h\to\gamma\gamma})$. In this paper, we study the possibility of measuring this production process at the LHeC  with the help of the photon PDFs published in recent years. This process has a clean final state without additional colored particles in the  detectable region other than the decay products of the Higgs. We compute the cross sections of all related processes and find that the production rate is at the same order as the neutral current WBF production of Higgs boson with missing forward jets. However,  a detailed phenomenological study of various Higgs decay channels shows that even in the most promising semi-leptonic $WW$ channel, the feasibility of identifying such photo-production is negative due to an irreducible photo-production of $W^{+}W^{-}$. 
\end{abstract}

\maketitle
\section{Introduction}

{ The discovery} of a 125~GeV Standard Model (SM) like Higgs boson by the ATLAS { and} CMS collaborations at the CERN Large Hadron Collider (LHC)~\cite{Aad:2012tfa,Chatrchyan:2012xdj}  has significantly improved our knowledge over { the} mechanism behind spontaneous electroweak gauge symmetry breaking.  
On the other hand, neither the mass of the Higgs boson nor the driving force of electroweak symmetry breaking is explained within the SM and these questions have motivated many attempts in extending { the} SM. Besides the direct search of such models of physics beyond SM (BSM), precision measurement of Higgs boson couplings and properties at the same time  plays an important role in testing various BSM physics models. For this purpose, proposals of future electron-positron  colliders, such as, FCC-ee, ILC and CEPC, have been widely discussed in the community.   Besides them, there also exists  a proposal for an $e$-$p$ collider known as { the Large Hadron-electron Collider (LHeC)}, planned to be constructed by adding { an} electron beam of 60-140~GeV to the current LHC~\cite{AbelleiraFernandez:2012cc}. LHeC was initially proposed as a TeV deep-inelastic scattering (DIS) facility with 60~GeV electron beam and 2~ab$^{-1}$ designed integrated luminosity to improve the measurement of parton distribution functions at large { Bjorken} $x$. By upgrading the electron beam energy and designed luminosity, it can potentially be converted into a ``Higgs factory''. The leading production  mode of Higgs boson at the LHeC is via charged-current  weak boson fusion (WBF) : $e+p \to \nu_e +h + j$. The tagged forward jet is { found to be} crucial  for suppressing the background in  a study on { the $h b\bar{b}$ coupling} at  the LHeC~\cite{Han:2009pe}.  Analogous to  the use of jet angular correlation in measuring the anomalous coupling in WBF at  the LHC~\cite{Hankele:2006ma}, the azimuthal angle between  the neutrino and forward jet $\Delta \phi_{\cancel{E}_{T}J}$   provides a sensitive probe of the anomalous $hWW$ coupling at the LHeC~\cite{Biswal:2012mp}.

{The diphoton} decay of Higgs boson played an important role in  { the Higgs discovery} at the LHC. At leading order, { the SM Higgs decay} into diphoton ($h\to \gamma\gamma$) arises from the $W$-loop and heavy quark  loop processes, where the $W$-loop dominates the decay width. Exotic particles from BSM models such as sfermions in supersymmetric models or charged Higgs { in the extended Higgs sector} may also contribute to the diphoton decay at one-loop level. Therefore, the loop-induced diphoton decay  provides a sensitive probe to physics at TeV scale. 
On the other hand, the photon-fusion production rate of Higgs boson is proportional to the Higgs diphoton decay width. {Therefore}, precision measurement of { the photon-fusion process could potentially be} complementary  to the diphoton decay measurement. 

Recent developments of photon PDF from CT14qed/LUXqed/NNPDF23qed~\cite{Ball:2013hta,Schmidt:2015zda,Manohar:2016nzj,Manohar:2017eqh}~have improved our framework for computing photon {initiated processes}. While the photon radiation off a point-like particle {e.g.} electron can be calculated explicitly or sometimes with Weizs\"{a}cker-Williams approximation, {photons from the proton} can arise from both elastic and inelastic processes. {The elastic channel contribution is from  photons directly radiated off a proton, while the inelastic channel is from those radiated off partons.} To verify the photon PDF {results, the} exclusive muon pair production via $\gamma\gamma \to \mu^+\mu^-$ has been measured for the first time by CMS~\cite{Chatrchyan:2011ci}. One complication {in measuring Higgs photo-production} at $e$-$p$ colliders is that the gluon fusion production of Higgs boson  contributes in $e$-$p$ collisions through $\gamma g$ scattering. In addition, the contribution from {WBF process with untagged forward partons} may not be neglected either. Hence, in order to explore the proton-production at the LHeC, the complete calculation and analysis of all these processes are important.

{The paper is organized as follows. In the next section, we discuss the {Higgs} photo-production {process and its cross section} at the LHeC. In section III, we will study the features of the signal, classify the background processes and {carry out a} phenomenological analysis {to separate} the signal and backgrounds. The results {will be} shown and briefly discussed in the last section.   }

\section{Higgs photo-production process at the LHeC}
\label{higgs_xsecs}
\begin{figure}[H]
\centering
	\subfloat[]{\includegraphics[scale=0.6]{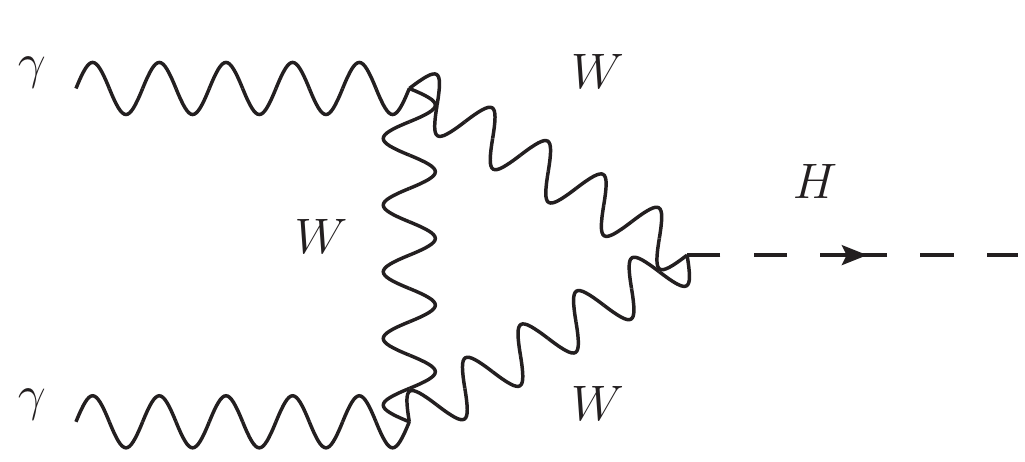}} \qquad
	\subfloat[]{\includegraphics[scale=0.6]{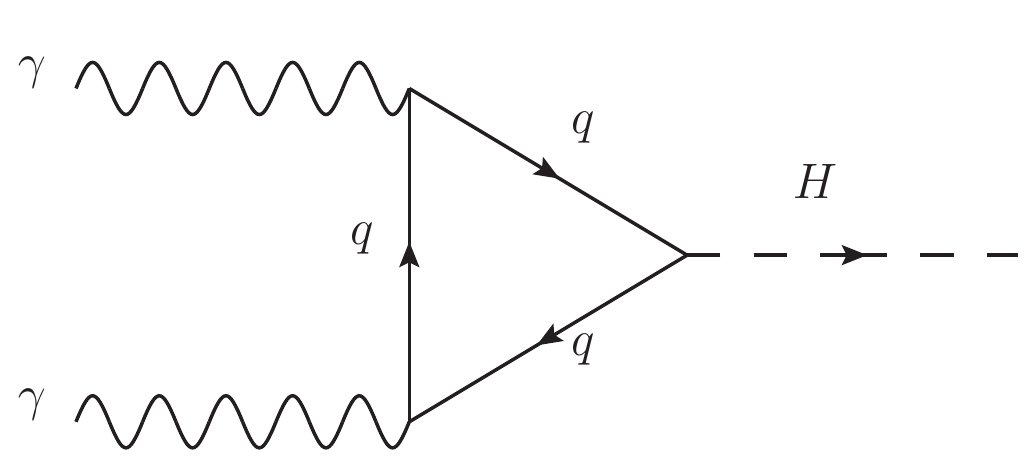}}\qquad
\caption{Representative Feynman diagrams of the Higgs photo-production. $q$ denotes any charged fermions. Top and bottom quark are dominant in the fermion loop.}
\label{fig_photo-production}
\end{figure}

The {Feynman diagrams of the} Higgs photo-production process through $W$ and charged fermion loops  are shown in Fig~\ref{fig_photo-production}.~The photons are radiated from the proton and electron beams. As the proton remnant has little recoil in the effective photon approximation~\cite{Budnev:1974de}, it moves in the very forward direction and {may even escape the detector}.~The process is therefore featured with its clean final states without additional colored particles. 
{The photo-production cross sections are summarised in Table~\ref{phpXsection} for two electron beam energies.~For comparison we list the result computed with {\tt NNPDF23\_nlo\_as\_0119\_qed}~\cite{Ball:2013hta}, and two more recent PDF sets {\tt CT14qed\_inc\_proton}~\cite{Schmidt:2015zda} and {\tt LUXqed17\_plus\_PDF4LHC15\_nnlo\_100}~\cite{Manohar:2016nzj,Manohar:2017eqh}, {obtaied from the LHAPDF6}~\cite{Buckley:2014ana}. These cross sections differ by a sizable amount
because the methods in determining these photon PDFs are quite different.  The LUXqed17 result is smaller than NNPDF23qed at
large momentum fraction $x$, while gets close when $x$ becomes smaller ($\sim \mathcal{O}(10^{-2})$). 
The CT14qed\_inc cross section is smaller than LUXqed17 in a more broad $x$ region.
These observations are in qualitative agreement with the behaviors of the photon PDFs shown in Fig.4 of the Ref.~\cite{Manohar:2016nzj}.}
\begin{table}[H]
\begin{center}
\begin{tabular}{|c|c|c|c|}
\hline
	Fig~\ref{fig_photo-production} & NNPDF23qed & CT14qed\_inc & LUXqed17 \\
\hline
	$E_e=60$~GeV & 0.52  & 0.39 & 0.46 \\
\hline
	$E_e=120$~GeV & 0.74   & 0.64 & 0.74 \\
\hline
\end{tabular}
\end{center}
	\caption{Cross sections in fb for processes represented by the diagrams in Fig.~\ref{fig_photo-production}, computed with three different PDF sets. The electron beam energy is $60/120$~GeV.}
\label{phpXsection}
\end{table}
{The results above are obtained with the full electron-photon splitting process considered. One could also carry out the calculation using Weizs\"{a}cker-Williams approximation~\cite{Budnev:1974de}.~This approach gives for instance, a cross section of 0.48 fb, for $E_e=60$~GeV, with NNPDF23qed. The accuracy of the approximation is satisfactory.}

\section{phenomenological analysis}
{\label{pheno}}
{In this section we study the possible background processes for the Higgs photo-production signal, and how to suppress them. There are two classes of backgrounds, those with and without a Higgs boson produced. The former ones will, of course, appear in all Higgs decay channels. This allows a study of the Higgs production subprocess without considering its decay, which will be done in section~\ref{bkg}. On the other hand, processes without an intermediate Higgs may also produce final states that mimic the Higgs decay product in various decay channels. However, the background contamination in this case will differ channel by channel. We shall discuss this class of backgrounds in section~\ref{kine} along with the simulation of the Higgs decays.}
\subsection{Forward electron tagging}
\label{fw-tagging}
In the Higgs photo-production {process, one expects to observe only the decay product of the Higgs in the central rapidity region.} The forward final state electron is usually {assumed} to escape from the detector because of its large pseudo-rapidity $\eta$. {However,} with the forward detector at the LHeC~\cite{AbelleiraFernandez:2012cc}, abundant forward electrons {from signal events} would be visible. In this case, the forward electron tagging {can also be} an important method to separate the photo-production from other processes. An ongoing study~\cite{Photo} shows that in photo-production processes nearly half of the electrons fall into the region $|\eta| \leq 5$.   We could do this tagging to remove all backgrounds with no scattered forward electrons in the final state, e.g.~charged current WBF.  The result of the forward electron tagging  for various backgrounds will be shown in our analysis below.

\subsection{{Backgrounds with a Higgs produced}}
\label{bkg}
\begin{figure}[H]
\centering
\subfloat[]{\includegraphics[scale=0.5]{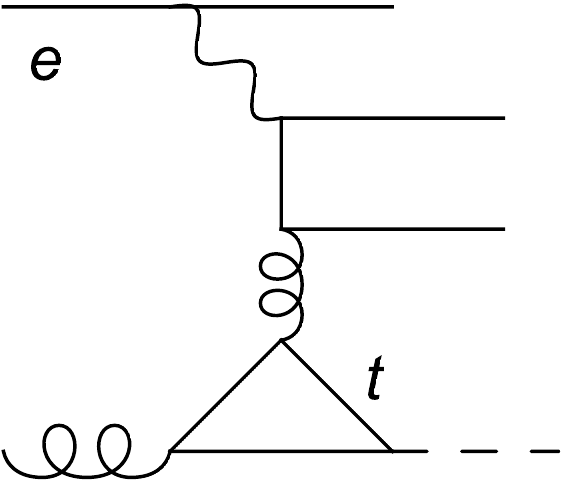}}\qquad
\subfloat[]{\includegraphics[scale=0.5]{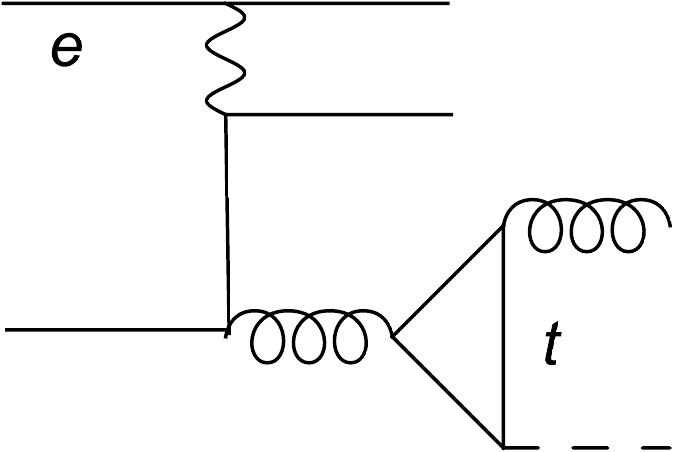}}\qquad
\subfloat[]{\includegraphics[scale=0.48]{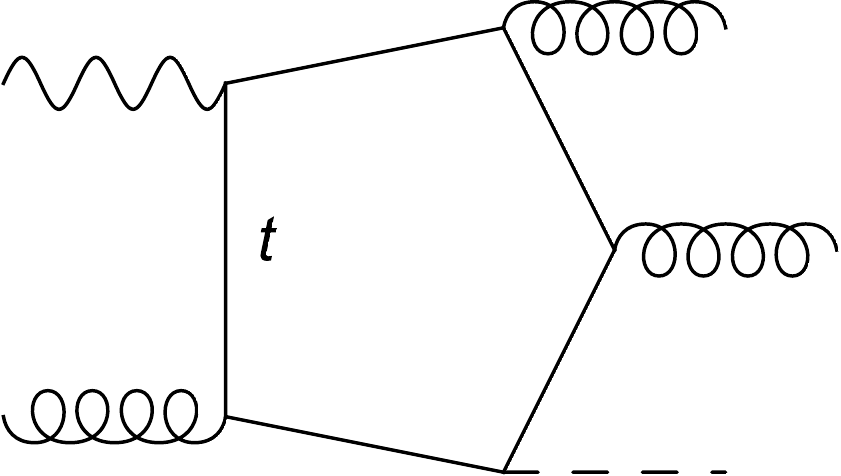}}\\
\subfloat[]{\includegraphics[scale=0.5]{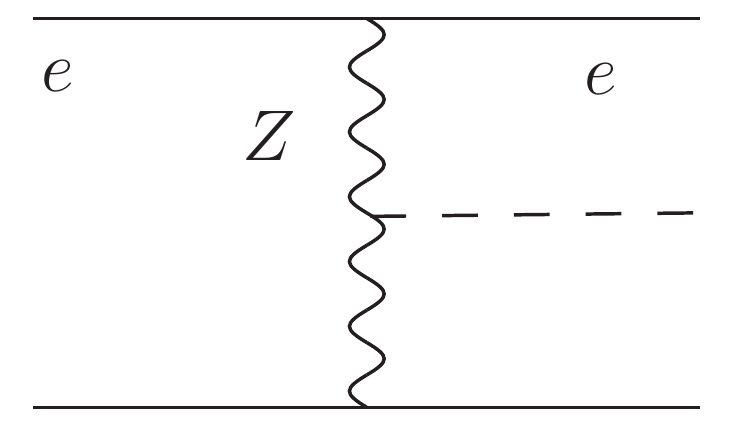}}\qquad
\subfloat[]{\includegraphics[scale=0.5]{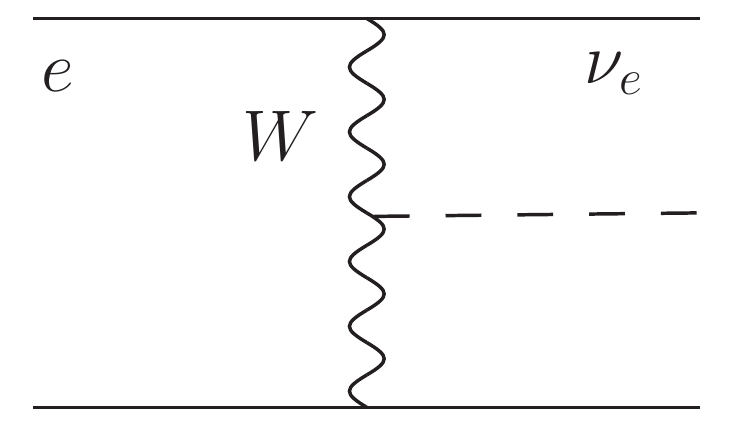}}
	\caption{{Representative Feynman diagrams for the {backgrounds with a Higgs produced}. The electron lines in (a) and (b) indicate that
	Weizs\"{a}cker-Williams approximation for photon radiations is not applicable in these cases and the electron-photon vertices are included explicitly
	in the calculation.}}
\label{fig_photo_diagrams}
\end{figure}

The Backgrounds with a Higgs produced have a common characteristic that their final state partons (other than those from the Higgs decay) are too soft or collinear to the beams to be tagged. Such processes, as shown in Fig~\ref{fig_photo_diagrams}, {can be} {irreducible backgrounds}. In particular, there are gluon initiated processes with more radiations. {With the increasing energy, the contribution from gluon PDF may quickly result in sizable background cross sections.}  {Therefore, these processes should not be simply ignored without knowing their contributions. In the following, we classify processes with their initial states and discuss the corresponding cross section calculations.}

\subsubsection{$\gamma g/\gamma q \to hX$}
\label{gamma_g_gamma_q}
The process represented by Fig.~\ref{fig_photo_diagrams} (a) also has a clean final state { in which hadrons with { small} rapidities come from the Higgs decay,}  despite the presence of additional radiation of the quark pair. {In fact, when the electron and quark masses $m_e$ and $m_q$ are much smaller 
than the center-of-mass energy $\sqrt{s}$,  the strongly ordered multiple splittings from Fig.~\ref{fig_photo_diagrams} (a)
give terms proportional to
\bea
\centering
\alpha_{EM}^2\alpha_S \bigl [\ln^3 \bigl (\frac{s}{m_e^2}\bigr )-\ln^3 \bigl (\frac{m_q^2}{m_e^2}\bigr )\bigr ]. 
\label{trilog}
\eea
These triple logarithms could substantially enhance the cross section in the region where the quark pair is collinear to the electron.}
Because of this  enhancement, the cross section from this channel receives large
contribution from the outgoing electron in the very forward region.~In principle, one could represent the
structure of the electron by parton distribution functions that evolve to account for the effects of the
large logarithms to all orders. { However, as we shall see, the cross section of this channel  
turns out to be quite small. We therefore ignore processes with more radiations and perform an order-of-magnitude estimation using
only the diagrams in e.g. Fig.~\ref{fig_photo_diagrams} (a), whose collinear
 singularities are cut off by the electron mass.}
 Also note that due to the multi-logarithmic
 structure of radiations,  Weizs\"{a}cker-Williams approximation cannot be used in this case, since additional
 collinear singularities exist in the hard scattering process even after the electron splitting vertex
 is factorized. We therefore include the electron line in Fig.~\ref{fig_photo_diagrams} {(a)} to indicate that the exact
 electron-photon vertex is included in the calculation. 
 The phase space integration of this channel is subject to large numerical uncertainty due to the rapid increase of
 the scattering amplitude in the collinear region.

Similar to the case in Fig.~\ref{fig_photo_diagrams} {(a)}, the quark radiation in Figs.~\ref{fig_photo_diagrams} {(b)} 
is also enhanced, but with more mild double logarithms of the form
\bea
\centering
\alpha_{EM}^2 \bigl [\ln^2 \bigl (\frac{s}{m_e^2}\bigr )-\ln^2 \bigl (\frac{m_q^2}{m_e^2}\bigr )\bigr ]. 
\label{trilog}
\eea
In contrast, the gluon radiations from Fig.~\ref{fig_photo_diagrams} {(c)} are not enhanced at large rapidities of the final gluons,
in which case { single logarithms $\alpha_{EM}\ln(s/m_e^2)$ appear as  prescribed by} Weizs\"{a}cker-Williams approximation.  
Note that one cannot invoke Furry's theorem to discard the pentagon graphs for two obvious reasons:
the number of $\gamma$ matrices from the vertices in the loop is even; and there is a mixing of QED and QCD vertices.
\begin{table}[H]
\begin{center}
\begin{tabular}{|c|c|c|c|}
 \hline
	$E_e=60$~GeV   &  Figs~\ref{fig_photo_diagrams}(a) &  2(b) & 2(c) \\\hline
NNPDF23qed    & $\sim$0.017 & $\sim7.5\times 10^{-4}$ & $1.9\times 10^{-4}$ \\\hline
CT14qed\_inc    & $\sim$0.016 & $\sim7.6\times 10^{-4}$ & $2.0\times 10^{-4}$ \\\hline
LUXqed17     & $\sim$0.017 & $\sim8.2\times 10^{-4}$ & $2.0\times 10^{-4}$  \\\hline
 \hline
	$E_e=120$~GeV   &  Figs~\ref{fig_photo_diagrams}(a) &  2(b) & 2(c) \\\hline
NNPDF23qed   & $\sim$0.038  & $\sim$0.0016  &  $7.8\times 10^{-4}$  \\\hline
CT14qed\_inc    & $\sim$0.035 & $\sim$0.0016 & $8.1\times 10^{-4}$ \\\hline
LUXqed17  & $\sim$0.034  & $\sim$0.0018  &  $8.1\times 10^{-4}$  \\\hline
\end{tabular}
\end{center}
	\caption{Cross sections in fb for processes represented by the diagrams in Fig.~\ref{fig_photo_diagrams}(a), (b), and (c), computed with three different PDF sets. The electron beam energy is $60/120$~GeV.}
\label{tableAgAq}
\end{table}
{The cross sections for $\gamma g$ and $\gamma q$ processes  are shown in Table~\ref{tableAgAq}, where the ``$\sim$'' indicates the number is presented with large uncertainty and serves only as an order-of-magnitude estimate, for the reason discussed above.
The three PDF sets give very close results. These cross sections are at most $\sim 5\%$ of that of the photo-production at the energies we are considering. Hence, we will neglect the contributions from the diagrams (a), (b) and (c) in Fig.\ref{fig_photo_diagrams}.}

\subsubsection{Weak boson fusion}
WBF processes as in Fig.~\ref{fig_photo_diagrams} (d) and (e) are the dominant for Higgs production at the LHeC with a large cross section of $\mathcal{O}(10^{2})$ fb. 
Therefore we expect some rate from the WBF processes without spectator partons (those not from Higgs decay) in the detectable region. To calculate the cross section we exclude the region $|\eta^{q}|\leq5$,\footnote[1]{We do not do this for $\gamma q$ and $\gamma g$ processes because their cross sections are negligible already.} where $q$ denotes the final state quark (not from Higgs decay).
 There is again little PDF dependence for the cross sections shown in Table~\ref{tableWBF}. The neutral current WBF cross sections are comparable to those of the photo-production, while the charged current ones are about an order of magnitude larger.   However, the charged current process can be eliminated completely by forward electron tagging as discussed in Sec.\ref{fw-tagging}.   Therefore, we will not consider  it in our following simulation. 
{
\begin{table}[H]
\begin{center}
\begin{tabular}{|c|c|c|c|c|c|c|}
 \hline
	$E_e=60$~GeV  &Fig~\ref{fig_photo_diagrams}(d), $|\eta^{q}|\geq5$ & Fig~\ref{fig_photo_diagrams}(e), $|\eta^{q}|\geq5$\\\hline
NNPDF23qed   & 0.68 & 4.8\\\hline
CT14qed\_inc   & 0.69 & 5.0\\\hline
LUXqed17   & 0.70 & 4.9 \\\hline
 \hline
	$E_e=120$~GeV & Fig~\ref{fig_photo_diagrams}(d), $|\eta^{q}|\geq5$ & Fig~\ref{fig_photo_diagrams}(e), $|\eta^{q}|\geq5$\\\hline
NNPDF23qed  &  1.08 & 7.5 \\\hline
CT14qed\_inc   & 1.09 & 7.6\\\hline
LUXqed17  & 1.10  & 7.6 \\\hline
\end{tabular}
\end{center}
	\caption{Cross sections in fb for processes represented by the diagrams in Fig.~\ref{fig_photo_diagrams}(d) and (e), computed with three different PDF sets. The electron beam energy is $60/120$~GeV.}
\label{tableWBF}
\end{table}
}
{Throughout this paper} we choose the renormalization and factorization scales to be the center-of-mass energy for the hard scattering processes.
The reduction of the pentagon loop integral in Fig.~\ref{fig_photo_diagrams} (c) requires extra numerical accuracy and is done with the help of Madloop program~\cite{Hirschi:2011pa} 
in {\it MadGraph5\_aMC@NLO}~\cite{Alwall:2014hca}. Other loop diagrams are calculated with FormCalc and LoopTools~\cite{Hahn:1998yk}. The Monte Carlo phase space integration is performed using VEGAS algorithm  implemented in CUBA library~\cite{Hahn:2004fe}.

\subsection{Simulation and selection cuts}
\label{kine}
In this part we explore the possibility of separating the Higgs photo-production---to be taken as our signal process---from other backgrounds with similar final states. The final state particles from the background can be produced from the Higgs decay {after the processes} shown in Fig.\ref{fig_photo_diagrams}. On the other hand, {there are backgrounds from $\gamma g$ and $\gamma \gamma$ scatterings {that can produce final states similar to those from various Higgs decay channels}. To separate all these backgrounds from the signal 
we make use of the kinematic features of three main decay channels ($h \to$$b\bar{b}$, $W^+W^-$ and $ZZ$). The background simulation is performed with {\it MadGraph5\_aMC@NLO}~\cite{Alwall:2014hca} at the parton level, {and we use {\it Pythia6.420} \cite{Sjostrand:2006za} and {\it Delphes3.3.0} \cite{deFavereau:2013fsa} for parton shower and detector simulations respectively.} The PDF set {\tt NNPDF23\_nlo\_as\_0119\_qed} is used in simulations of both the signal and backgrounds, which includes {both} elastic and inelastic photon information~\cite{Ball:2013hta}. {As is done in Sec.\ref{higgs_xsecs},} we choose two benchmark electron beam energies at $60$~GeV and $120$~GeV with $7~$TeV proton beam to see whether increasing the electron beam energy helps to improve the Higgs production measurement. The basic cuts on final states $p_T$, $\eta$ and $\Delta R$ are applied as
\bea
\centering
p^{\ell}_{T}\geq5~\text{GeV}&,&p^{j}_{T}\geq20~\text{GeV},\nonumber \\
|\eta^{\ell}|\leq5&,&|\eta^{j}|\leq5,\nonumber \\
\Delta R_{\ell\ell}\geq0.4&,&\Delta R_{jj}\geq0.4, 
\label{basiccut}
\eea
{where $l$ and $j$ denote a final state lepton and parton respectively.}

\subsubsection{$b\bar{b}$}
The standard $h\to b\bar{b}$ search at the LHeC uses forward jet tagging to improve the signal-to-background rate, which {makes possible the} bottom Yukawa measurement~\cite{Han:2009pe}. {In our search the main background {processes are $\gamma g \to b\bar{b}$ and $\gamma g \to c\bar{c}$, with $c$ misidentified as $b$ in the detector}}. The large {difference} between gluon and photon PDFs in the proton makes the background cross-section orders larger than that of the signal. In order to pick out the signal events, the $b$-jet transverse momentum could be used because {the $b$ quarks} from Higgs decay are more boosted. The $b\bar{b}$ invariant mass is another {discriminative kinematic variable} and only events with $m_{b\bar{b}}$ around $m_h$ should be kept. { The $b\bar{b}$ search efficiency also depends {heavily} on the $b$-tagging ($\epsilon_{b}$) and $c$ faking ($\epsilon_{c \to b}$) efficiencies of the detector. Here we assumed $\epsilon_{b}=70\%$ and $\epsilon_{c\to b}=20\%$~\cite{Lapertosa:2016zpo}.} The signal production cross section is about $0.148$~fb {for} $E_e=60$~GeV and $0.211$~fb {for} $E_e=120$~GeV with $Br(h \to b \bar{b}) \approx 58\%$. {However, the $b\bar{b}$ background {cross sections for the two energies} are $92.1$~pb and $151.2$~pb, and the misidentified $c\bar{c}$ backgrounds are $31.7$~pb and $52.1$~pb.} Even after we apply kinematic cuts on the invariant mass of two $b$ jets --- $m_{bb}$, the larger transverse momentum of the two $b$ { jets} ---$^{1}p^{b}_T$, and $\Delta\eta=|^{1}\eta^{b} -^{2}\eta^{b}| $ as shown in Fig.\ref{bbchannel}, about $6\%$ of the total background events could survive. 
This is still several orders larger than the raw signal events. For this reason, it's extremely challenging to use $b\bar{b}$ channel for this measurement. 

\begin{figure}[H]
\centering
	\subfloat[ $^{1}p^{b}_T$ distribution]{\includegraphics[scale=0.36]{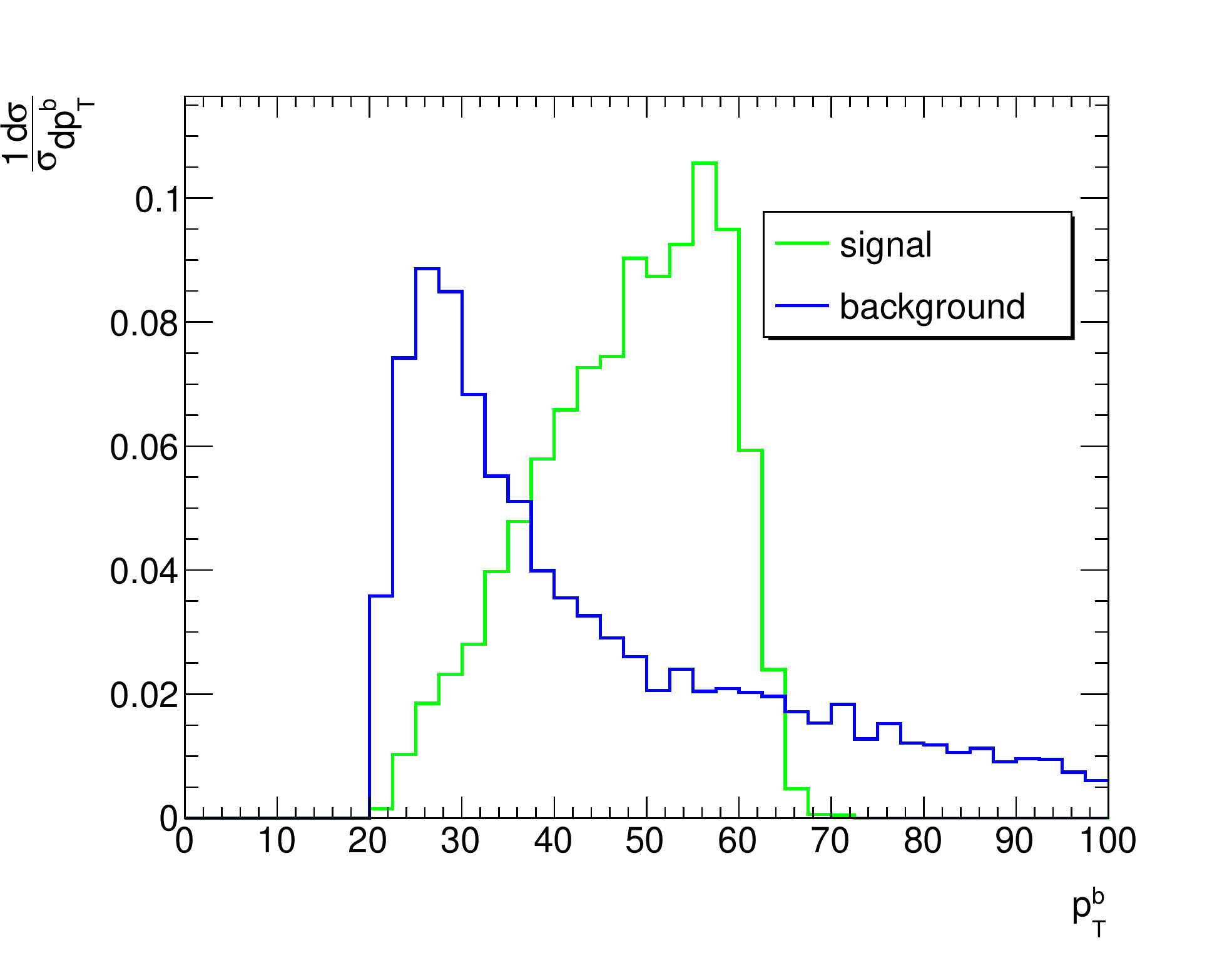}}\qquad
	\subfloat[ $\Delta\eta$ distribution]{\includegraphics[scale=0.36]{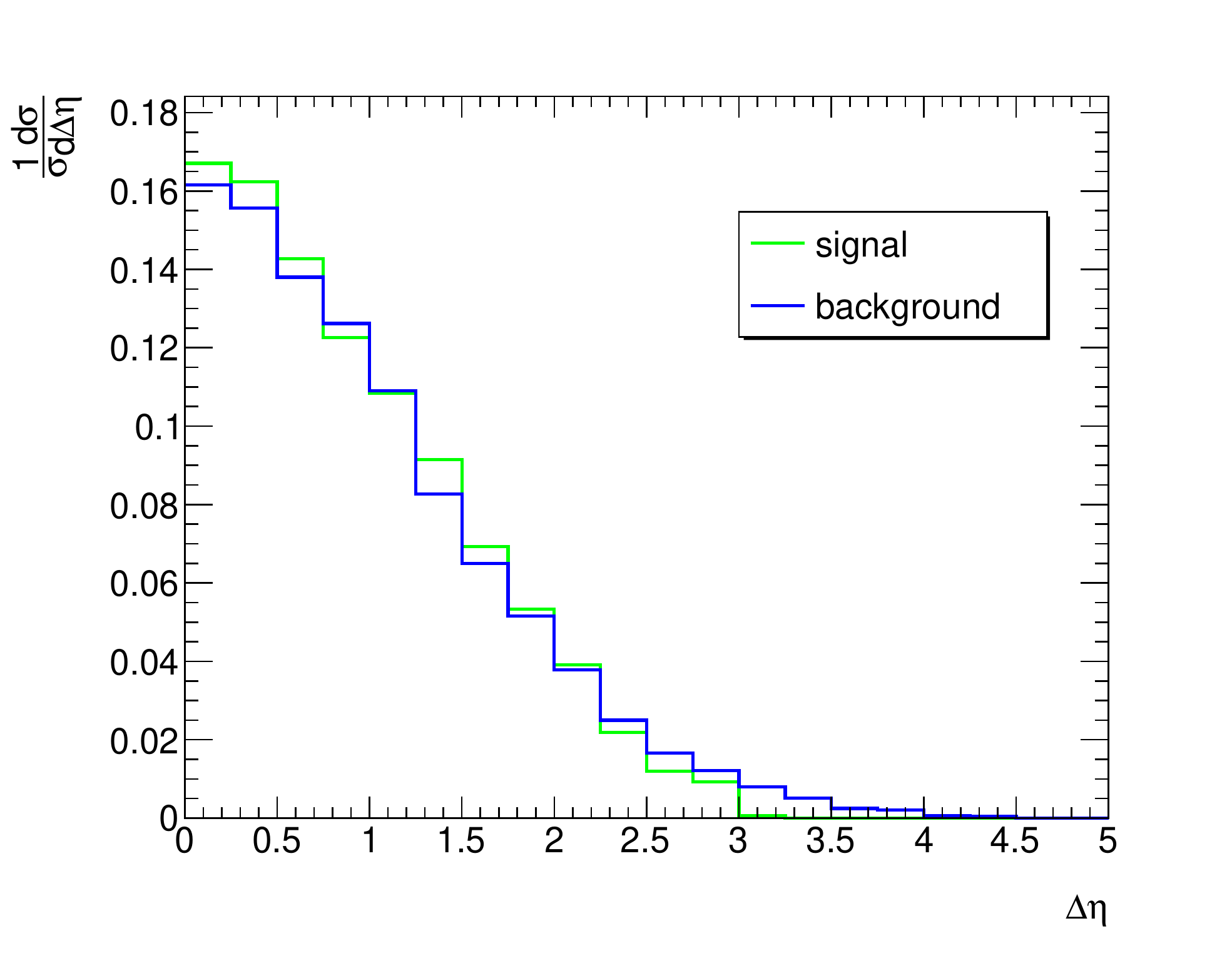}}\\
	\subfloat[ $m_{bb}$ distribution]{\includegraphics[scale=0.36]{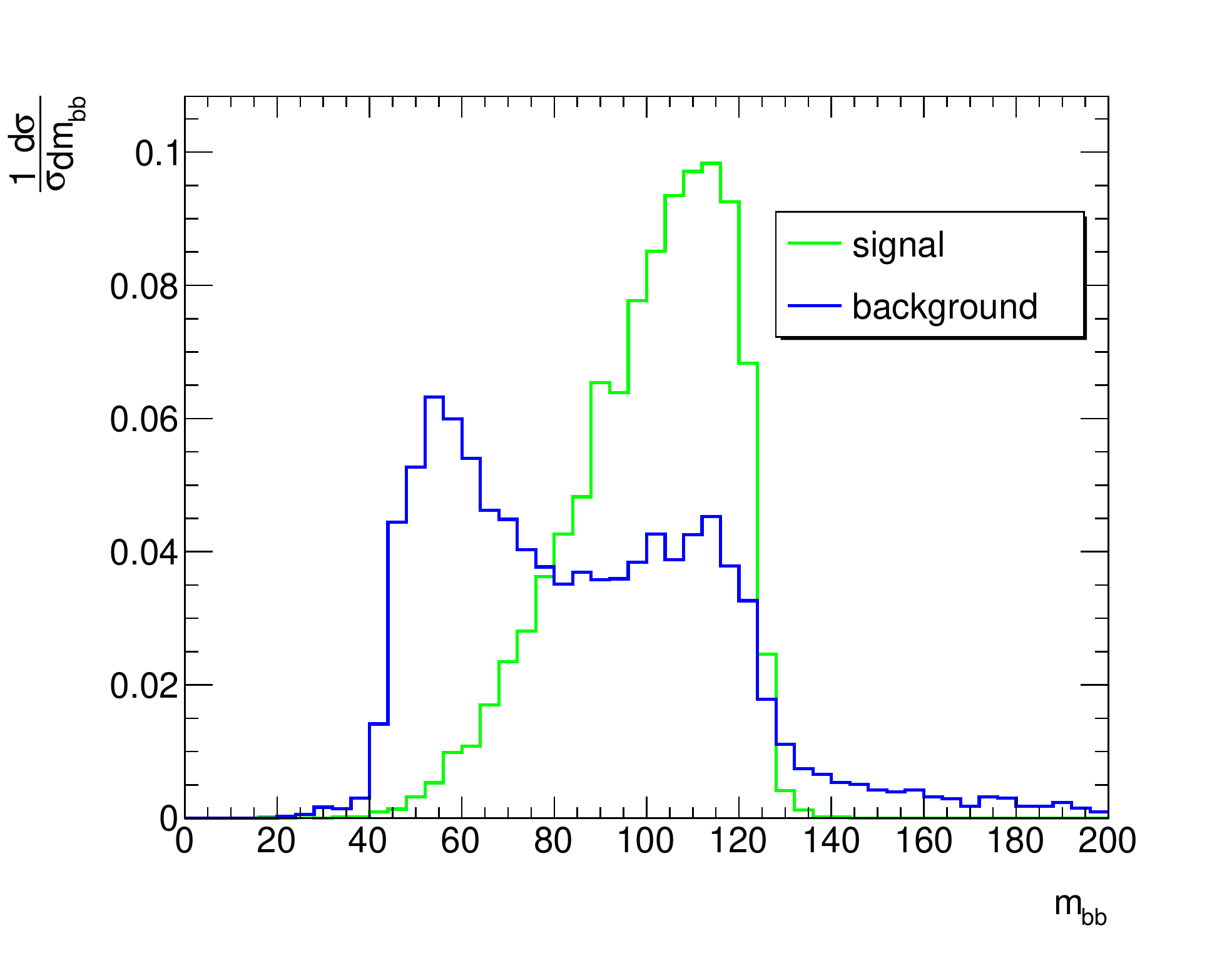}}

\caption{{ (a), (b) and (c) are distributions of $^{1}p^{b}_T$, $\Delta\eta$ and $m_{bb}$ for the signal (green) and background (blue), reconstructed from events after basic cuts.} The electron beam energy is $E_e=60~$GeV.}
\label{bbchannel}
\end{figure}

\subsubsection{$W^+W^-$}
The $W^+W^-$ channel is the secondary {Higgs} decay channel with $Br(h\to W^{+}W^{-}) \approx 21.5\%$. {In this study, we consider the semi-leptonic $W$ decay, in which one $W$ boson decays to $\ell \nu_{\ell}$ ($\ell = e,~\mu$) and the other to hadronic final states (two jets $jj$). {  Note that in the pure leptonic $W$ decay channel there is huge background from $\gamma g$ and $\gamma \gamma$ scatterings, while the corresponding background for the semi-leptonic $W$ decay is much more moderate. There can be the background from $\gamma\gamma\to \tau^+\tau^-$ with the semi-leptonic decay of the $\tau$'s.    However, the low invariant mass of the two jets from a $\tau$ decay can be easily distinguished from that in the decay of a $W$. Hence, we shall neglect the contamination from the $\tau$ decay. The main background will be from $\gamma\gamma\to W^+W^-$ with the semi-leptonic $W$ decay.  There is also the background from $\gamma g\to qq^{\prime}\ell\nu_{\ell}$, with two quarks $q$ and $q^\prime$, a lepton $\ell$ and a neutrino $\nu_{\ell}$ produced. As discussed before, the neutral current WBF process with forward spectator partons will be another background.} We present the analysis of this channel in two methods. 

{The signal and background processes all produce neutrinos that lead to $\slashed{E}_T$.}Though it's very difficult to reconstruct invariant {masses involving neutrinos}, kinematic methods for the intermediate-mass Higgs boson search could be applied in this case~\cite{Barger:1990mn}.~The {$jj\ell\slashed{E}_T$} system transverse mass could be reconstructed with the transverse {momenta} of the jets, leptons and the missing objects as 
\bea
m^2_{T} &\equiv& (E^{jj\ell}_T+\slashed{E}_T)^2-|{{ \vec{p}}}^{jj\ell}_T+\slashed{\vec{p}}_T|^2 \nonumber \\
&=&(\sqrt{|{  \vec{p}}_T^{j}+{ \vec{p}}_T^{j}+{\vec{p}}_T^{\ell}|^2+m^2_{jj\ell}}+|\slashed{\vec{p}}_T|)^2-|{\vec{p}}^{j}_T+{\vec{p}}^{j}_T+{\vec{p}}^{\ell}_T+\slashed{\vec{p}}_T|^2 
\label{mt}
\eea
The signal transverse mass distribution has an upper bound at $m_h$, while the background distribution is rather flat and much {larger}.~Fig.\ref{fig_leptonicW} (a) shows the distributions of {$jj\ell\slashed{E}_T$} system transverse mass. It's reasonable to set upper bounds for the two reconstructed observables to cut the background. As the signal lepton and  two central jets are from the Higgs cascade decay, the pseudo-rapidity difference $\Delta\eta$ between them should be smaller than backgrounds.~One can see {in Fig.\ref{fig_leptonicW} (b), (c) and (d) that} the distributions of the pseudo-rapidity difference, the missing transverse momentum and the final lepton transverse momentum could be effectively used to distinguish the signal and background. The kinematic distributions are not significantly dependent on the electron beam energy, so we only plot the $E_{e}=60~\text{GeV}$ case.
\begin{figure}[H]
\centering
\subfloat[$m_{T}$ distribution]{\includegraphics[scale=0.25]{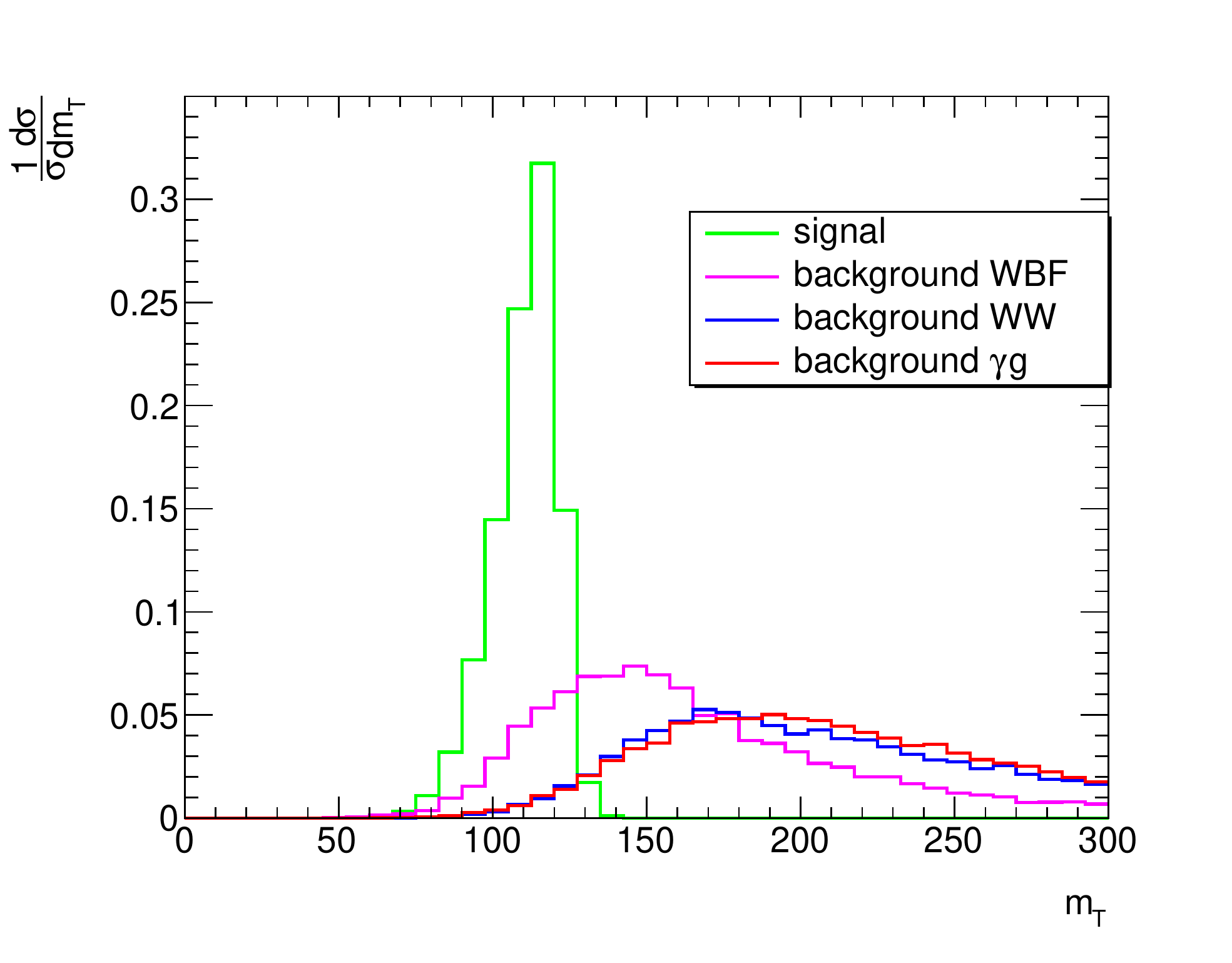}}
\subfloat[$\Delta\eta$ distribution]{\includegraphics[scale=0.25]{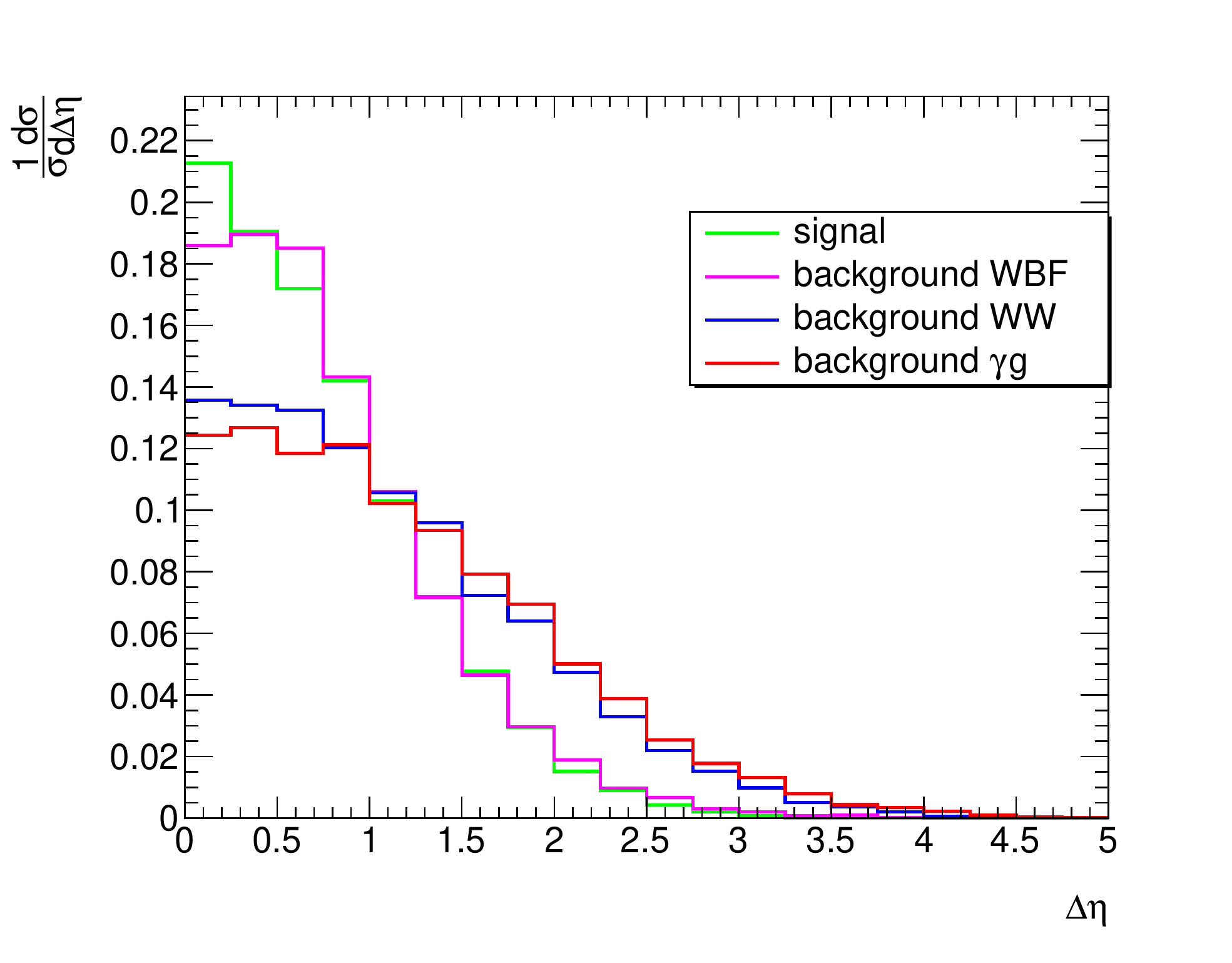}}
\subfloat[$m_{jj}$ distribution]{\includegraphics[scale=0.25]{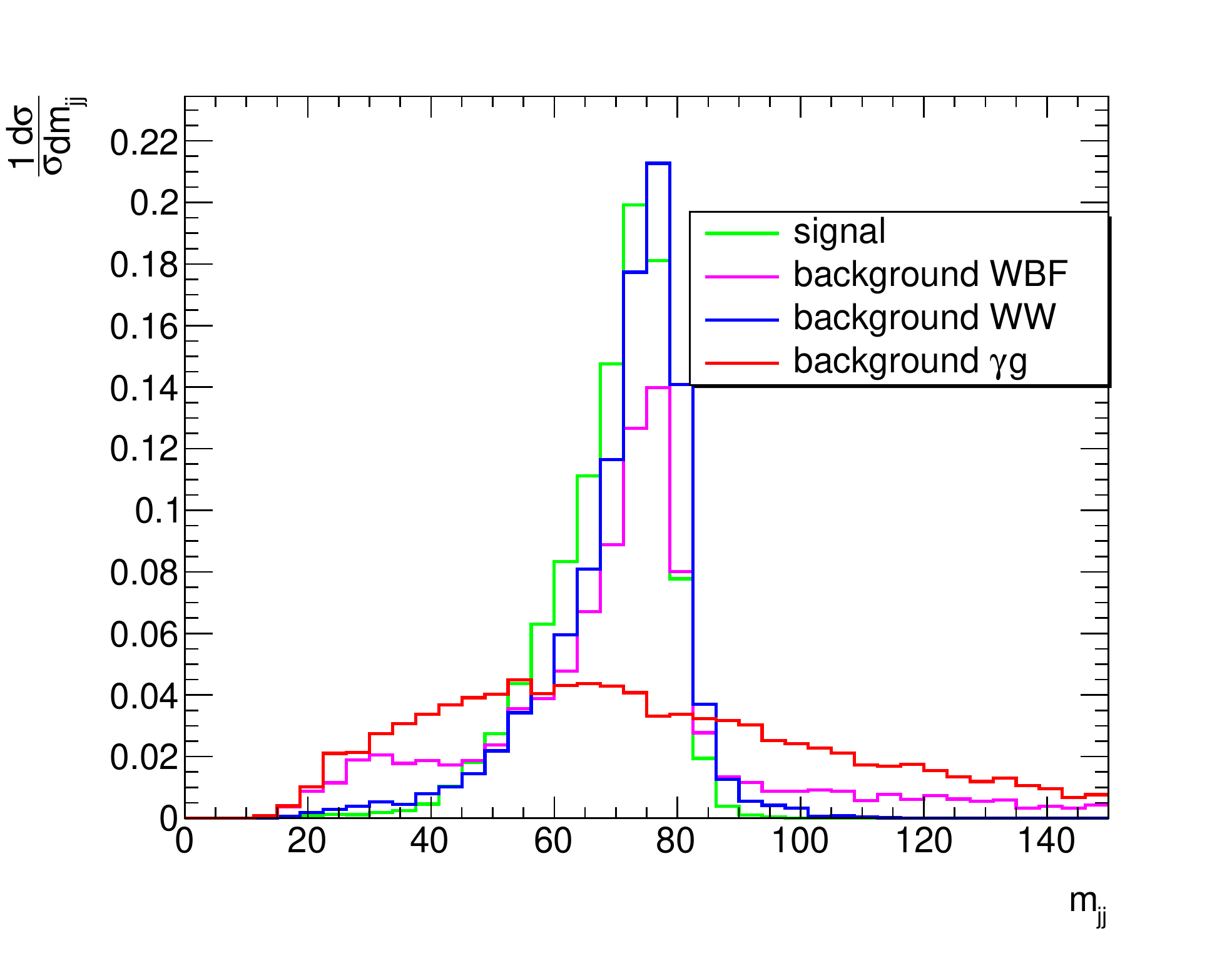}}\\
\subfloat[$\slashed{E}_T$ distribution]{\includegraphics[scale=0.25]{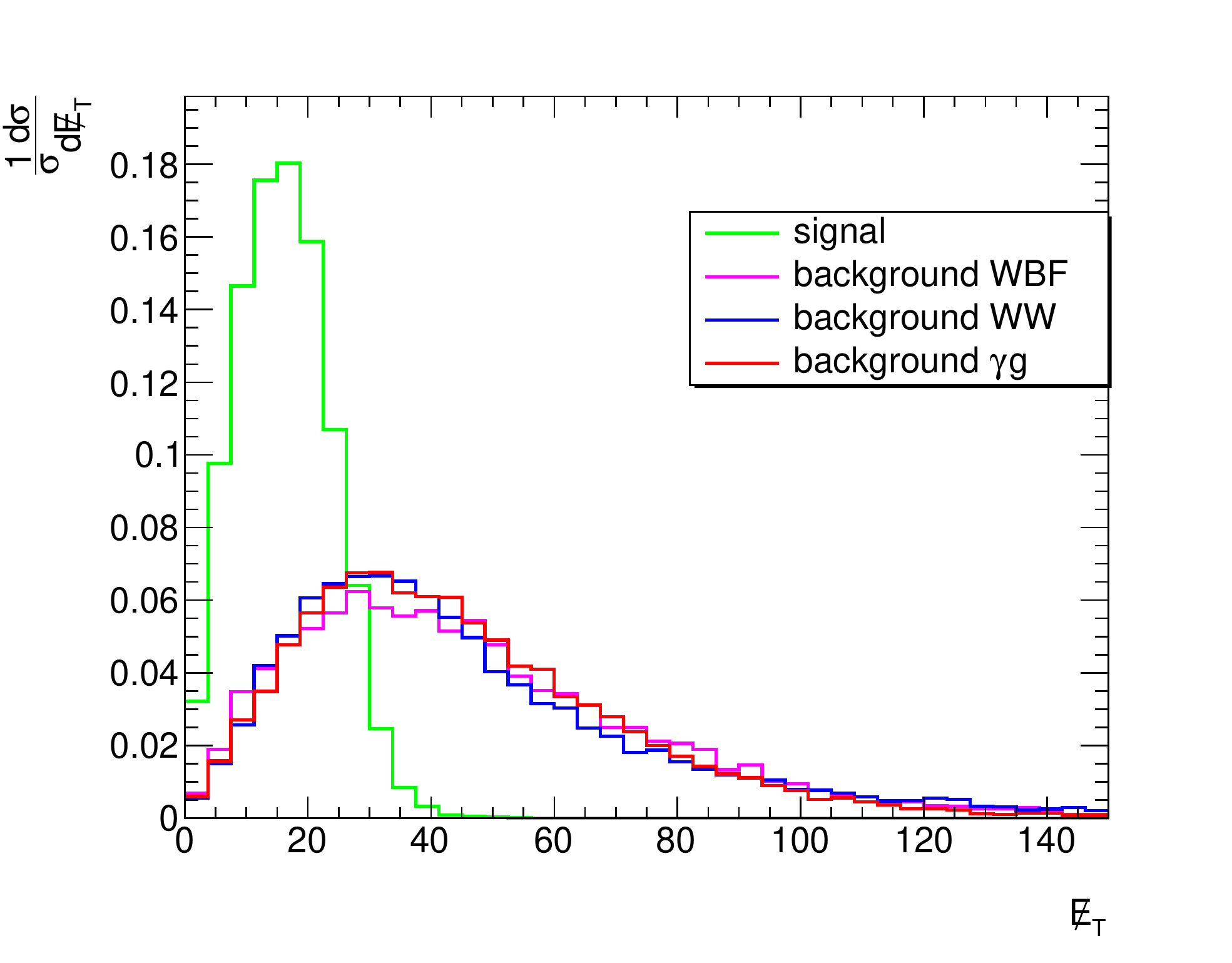}}\qquad
\subfloat[$p_{T}^{\ell}$ distribution]{\includegraphics[scale=0.25]{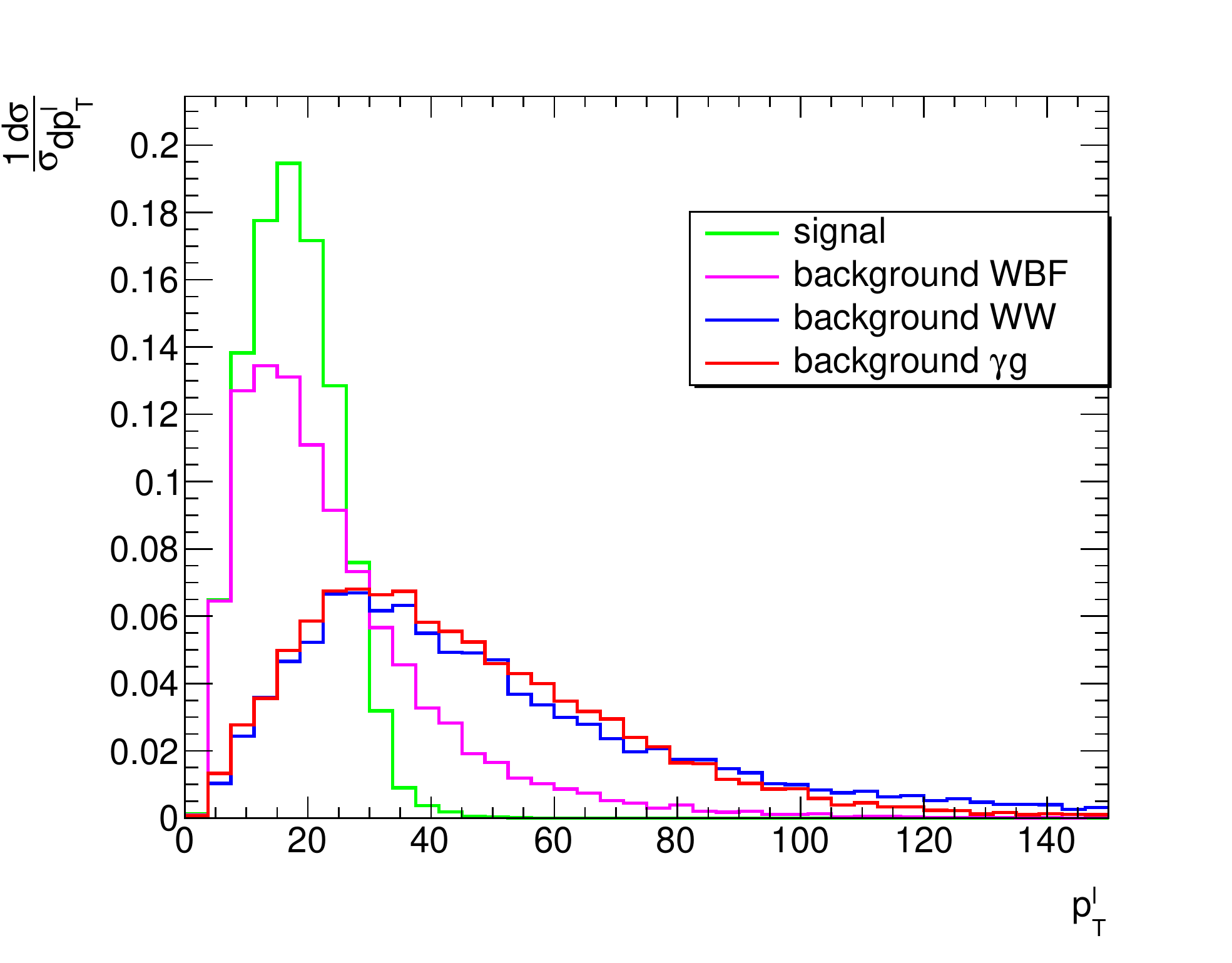}}
	\caption{ (a) (b) and (c) are distributions of $m_T$,  $\Delta\eta$, and $m_{jj}$ for the signal (green) and background (magenta, blue and red), reconstructed from events after basic cuts. Distributions of $\slashed{E}_T$ and {$p_{T}^{\ell}$} are shown in (d) and (e). The electron beam energy is $E_e=60~$GeV.}
\label{fig_leptonicW}
\end{figure}

According to the discussion, we require  $\slashed{E}_T\leq 30$~GeV, $p_T^{\ell}\leq 30$~GeV, $\Delta{\eta}\leq 2.5$,  $45~\text{GeV}\leq m_{jj}\leq 85$~GeV and $m_{T}\leq 125$~GeV. The efficiencies of signal and background events after all these cuts are listed in Table.\ref{tableww}.  Tagging the forward electron  removes the charged current WBF background completely, but it also  significantly reduces the signal, which makes the measurement  more difficult. Whether we perform the electron tagging or not, the signal is overwhelmed by the background, even though the kinematic cuts can reduce the background by about two orders.

\begin{table}[H]
\begin{center}
\resizebox{\textwidth}{30mm}{
\begin{tabular}{|c|c|c|c|c|c|c|c|}
 \hline
$E_e=60$~GeV & Cross Section ($fb$) & forward electron tagging & $p_T^{\ell}\leq 30$~GeV& $\slashed{E}_T\leq 30$~GeV  & $|\Delta{\eta}|\leq 2.5$ & $m_{T}\leq 125$~GeV & $45~\text{GeV}\leq m_{jj}\leq 85$~GeV   \\
\hline
Signal & $3.3\times 10^{-2}$ & 0.48 & 0.46 & 0.44 & 0.44 & 0.42 & 0.41 \\
\hline
Background WBF& $4.3\times 10^{-2}$ & 0.55 & 0.40 & 0.13 & 0.12 & 0.071 & 0.061  \\
\hline
Background $W^{+}W^{-}$ & $53.9$ & 0.50 & 0.15 & 0.026 & 0.021 & 0.011 & 0.011  \\
\hline
Background $\gamma g$ & $12.9$ & 0.50 & 0.16 & 0.025 & 0.020 & 0.0078 & 0.0064  \\
\hline
\hline
$E_e=120$~GeV & Cross Section ($fb$) & forward electron tagging & $p_T^{\ell}\leq 30$~GeV& $\slashed{E}_T\leq 30$~GeV  & $|\Delta{\eta}|\leq 2.5$ & $m_{T}\leq 125$~GeV & $45~\text{GeV}\leq m_{jj}\leq 85$~GeV   \\
\hline
Signal & $4.7\times 10^{-2}$ & 0.48 & 0.46 & 0.44 & 0.44 & 0.42 & 0.42  \\
\hline
Background WBF& $6.9\times 10^{-2}$ & 0.55 & 0.39 & 0.15 & 0.13 & 0.071 & 0.056 \\
\hline
Background $W^{+}W^{-}$ & $93.1$ & 0.50 & 0.15 & 0.026 & 0.020 & 0.010 & 0.010  \\
\hline
Background $\gamma g$ & $32.9$ & 0.50 & 0.16 & 0.025 & 0.019 & 0.0078 & 0.0058  \\
\hline
\end{tabular}
}
\end{center}
	\caption{The second column gives the cross sections of the signal and background after basic cuts. Other columns are cut efficiencies after corresponding kinematic cuts. {  Background WBF denotes the background  from the neutral current WBF process in Fig.\ref{fig_photo_diagrams}. } The electron beam energy is $60/120$ GeV.}
\label{tableww}
\end{table}

 The cut efficiencies above can be improved by performing a multivariate analysis with boosted decision trees (BDT). We carry out this analysis in the TMVA~\cite{Hocker:2007ht} framework. Our initial selection of events imposes the same basic cuts as above, and requires at least one lepton $\ell$ and two jets $jj$ in the final states. Then the following kinematic variables are used to train the BDT for the semi-leptonic final states. They are the $p_T$, $\phi$, and $E$  of the leading jet, sub-leading jet and leptons;  $\Delta \eta^{jj}$ and $\Delta \eta^{j\ell}$; $\slashed{E}$, $m_{jj}$, and $m_{jj\ell}$.   The BDT output and receiver operator characteristic (ROC) curves  in Fig.~\ref{fig_BDT_W} signify an improved separation of the signal and background. The resulting signal significances are 1.3$\sigma$ and 1.2$\sigma$, respectively for 60 GeV and 120 GeV electron beams.

\begin{figure}[H]
\centering
\subfloat{\includegraphics[scale=0.30]{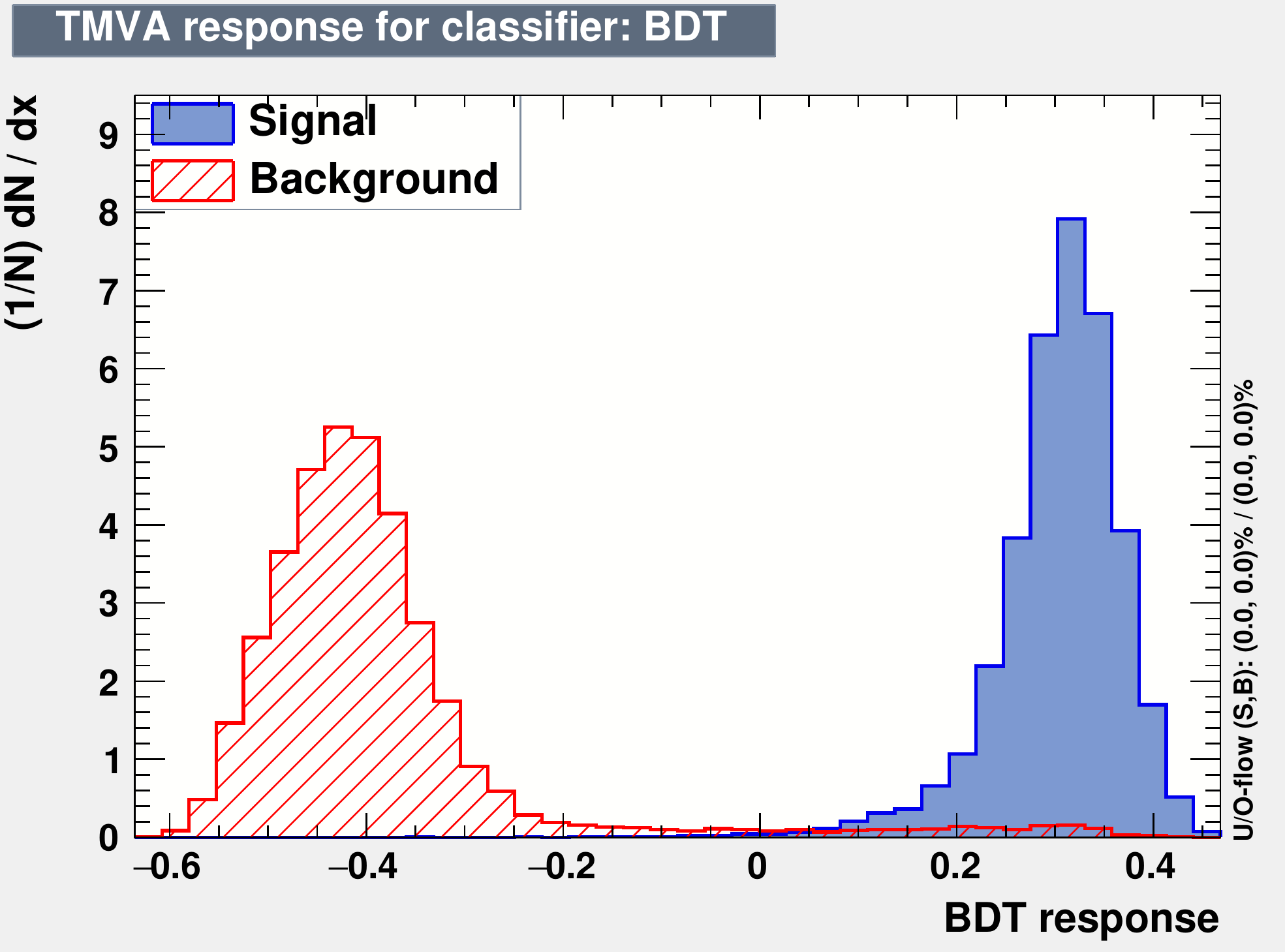}}\qquad
\subfloat{\includegraphics[scale=0.30]{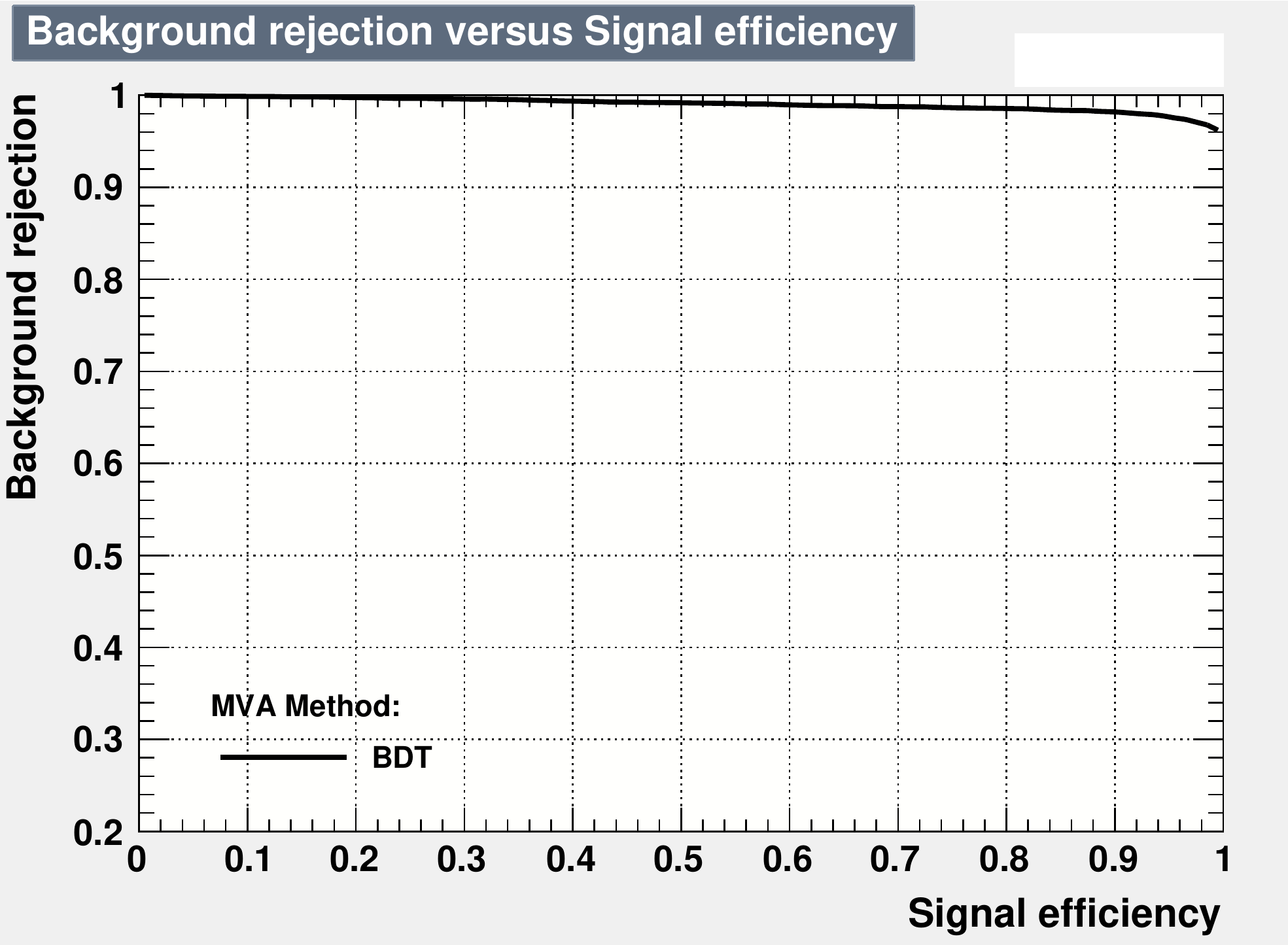}}\\
\subfloat{\includegraphics[scale=0.30]{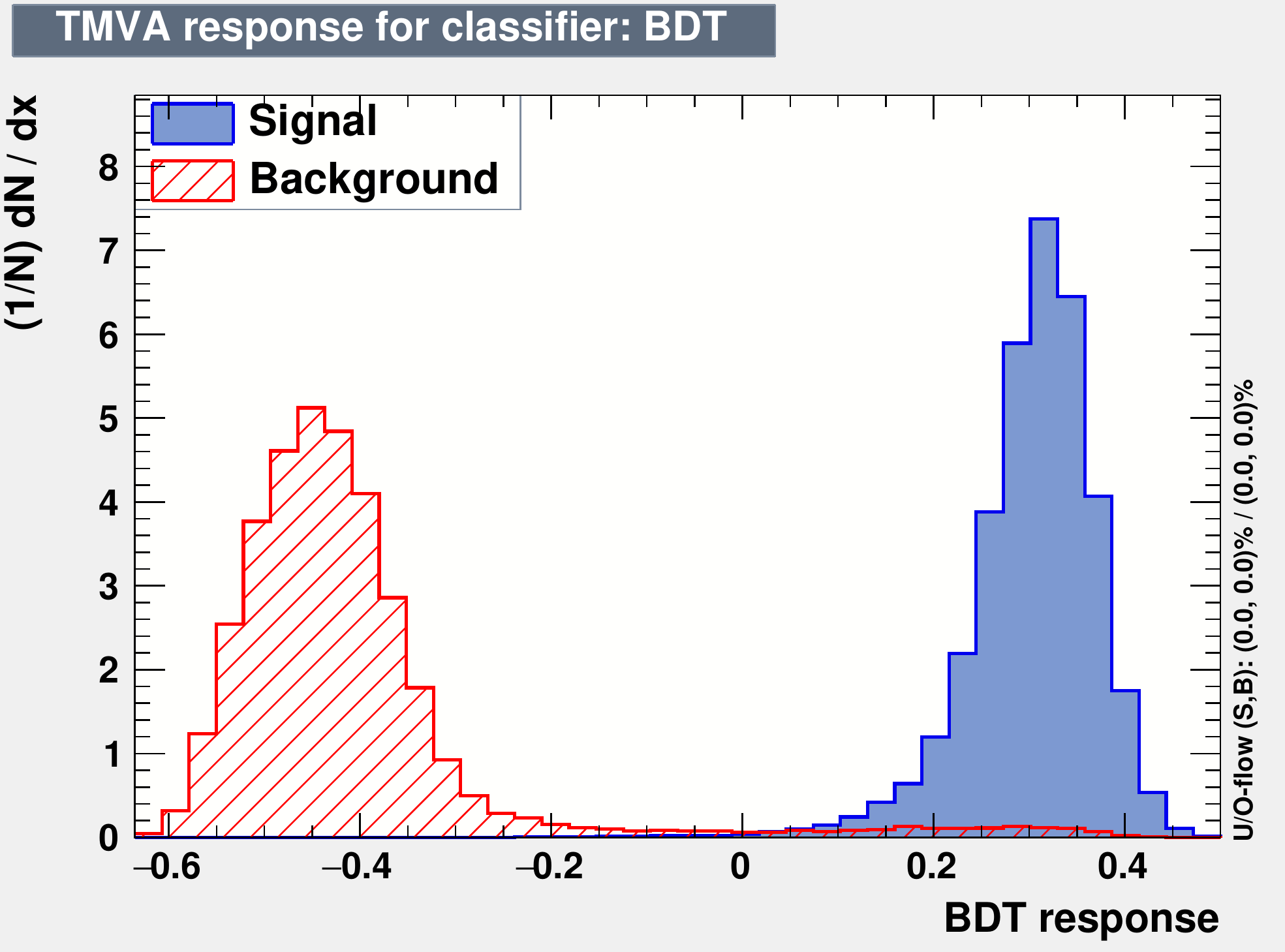}}\qquad
\subfloat{\includegraphics[scale=0.30]{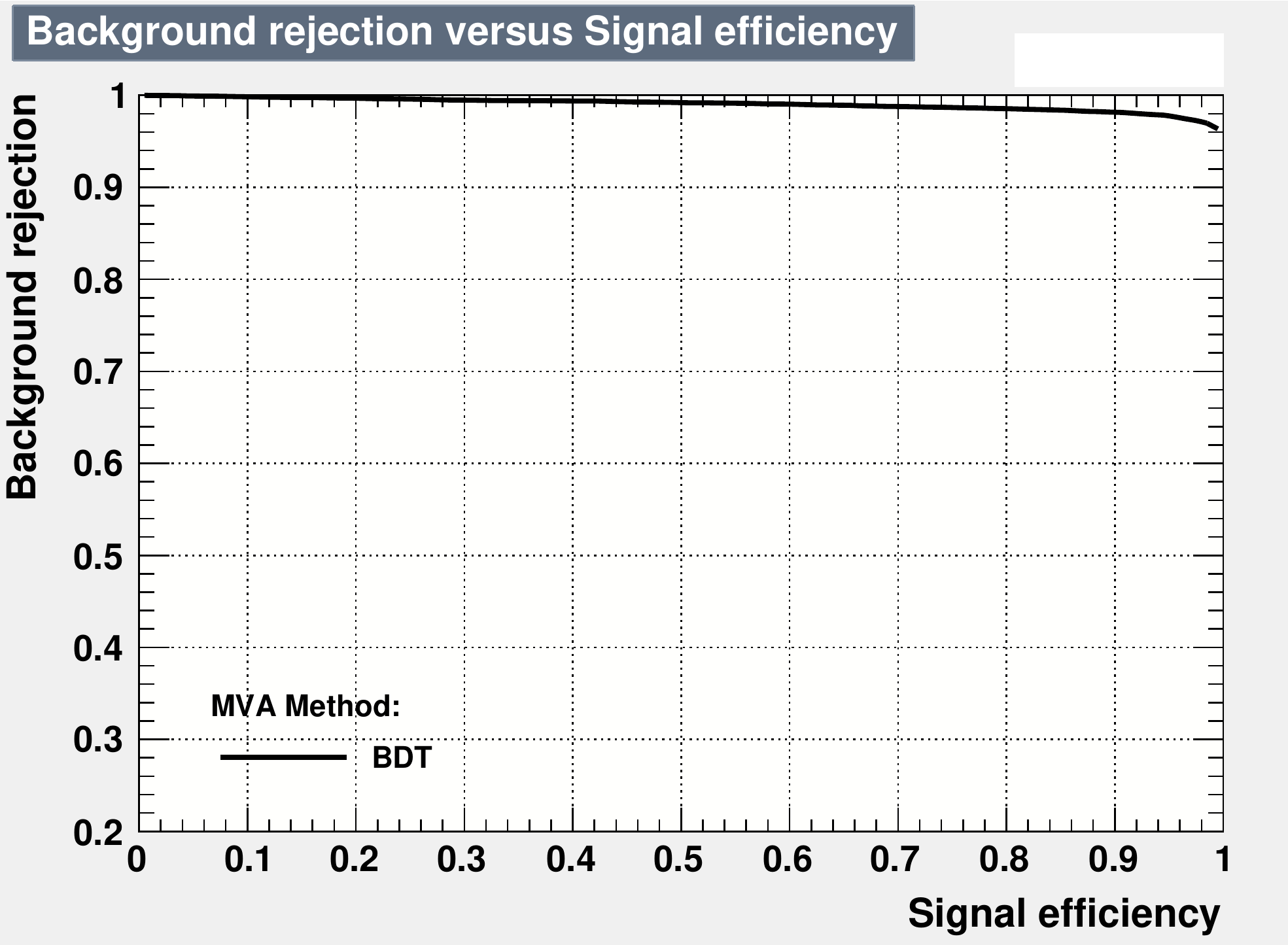}}
	\caption{{ Upper panels: Normalized BDT output (left) and receiver operator characteristic (ROC) curve of the BDT (right)} for the electron beam energy $E_e=60~$GeV. Lower panels: same as the upper panels but with $E_e=120~$GeV.}
\label{fig_BDT_W}
\end{figure}

\subsubsection{$ZZ$}
{The {$ZZ$-to-}four-lepton channel plays an important role in the discovery of Higgs boson and the measurement of its properties.~{ Here we  { focus on the muonic $Z$ decay} to extract the Higgs photo-production process. This helps to {exclude} huge QCD backgrounds and avoid the smear  { from the scattered electron.}}}~{There are two same-flavor and opposite-sign(SFOS) lepton pairs from $ZZ$ decay, which leads to the invariant mass of one SFOS pair $m_{\ell\bar{\ell}}$ around $m_{Z}$, and the total invariant mass $m_{4\ell}$ close to $m_h$}. At the meantime, the rapidity difference between the paired leptons should be small. After the lepton basic cut, the two SFOS leptons with invariant mass $^{1}m_{\ell\bar{\ell}}$ closest to the $Z$-boson mass are chosen as the first lepton pair. The requirement on $^{1}m_{\ell\bar{\ell}}$ is {$60\leq$$^{1}m_{\ell\bar{\ell}}\leq 95$}~GeV. The two remaining leptons form the second lepton pair, whose invariant mass should satisfy $^{2}m_{\ell\bar{\ell}}\leq60~$GeV. Since this process starts with an ergodic pairing, the background lepton-pair mass could possibly mimic the signal with a slightly broader distribution around $m_Z$. Even so, the $m_{4\ell}$ cut could still {eliminate} most of the background events because the background distribution is extremely flat, as shown in Fig.\ref{fig_4l}. Further kinematic requirements on lepton transverse {momenta} $^{1,2}p_T^{\ell}$ and pseudorapidity difference $^{1,2}\Delta\eta$ in the first and second lepton pairs are also used.
 Although with the kinematic cuts we could reduce the background to {less than} $1\%$ while {keep about} $70\%$ of the signal events, the huge difference between  signal and background cross sections makes it impossible to perform the measurement in this channel.

\begin{figure}[ht]
\centering
\includegraphics[scale=0.35]{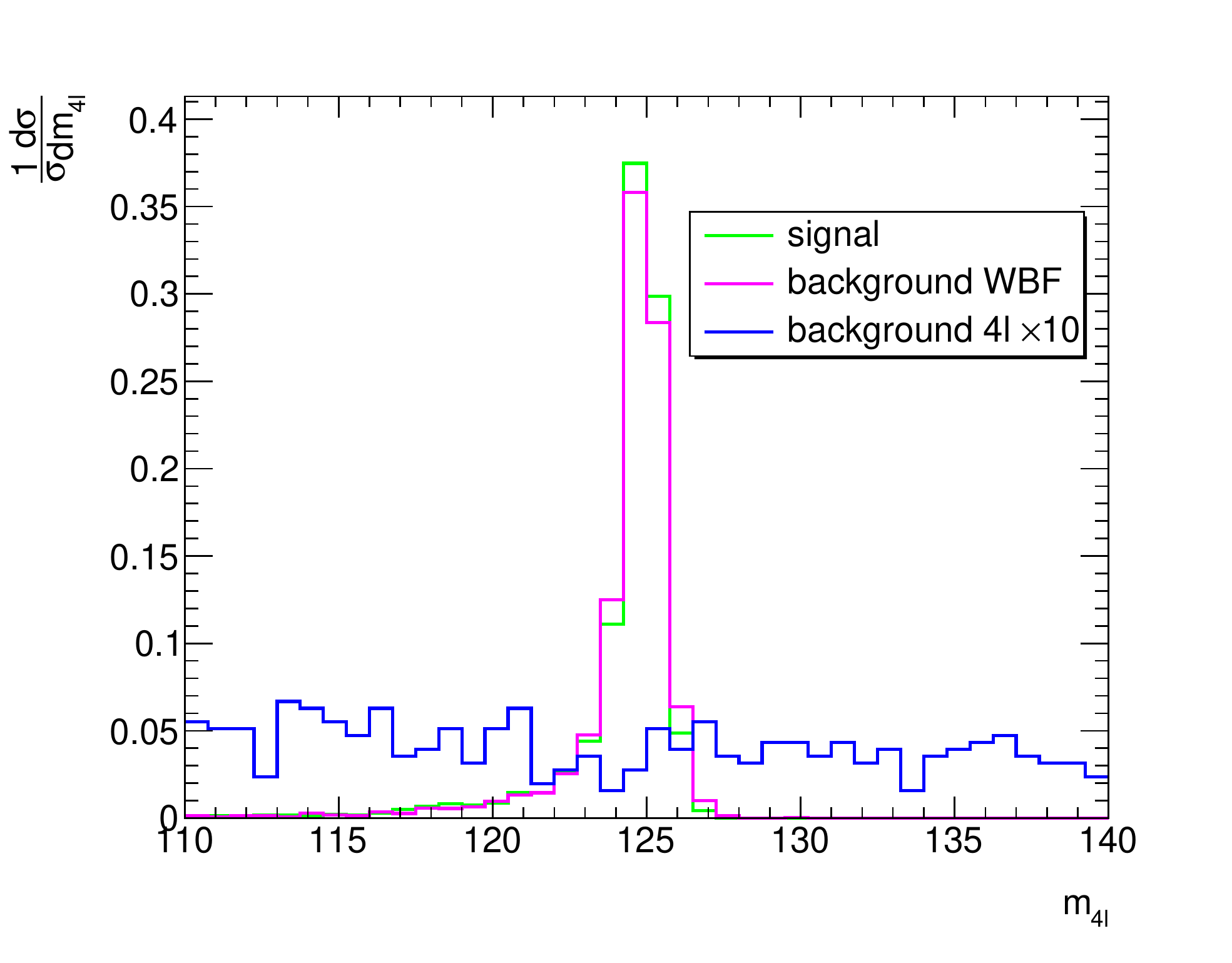}
	\caption{$m_{4\ell}$ distributions of the signal (green), WBF background (magenta) and tenfold scaled $W^{+}W^{-}$ background (blue).}
\label{fig_4l}
\end{figure}

\section{Results and Conclusion}
In this paper, we study the photo-production of the Higgs boson and related contaminating processes at the LHeC. {The cross sections are computed for all relevant processes.} We find that the photo-production is  overshadowed by other { Higgs production} processes whose spectator partons are  collinear { to one of the incoming beams} and cannot be tagged. 
 
	{ For three dominant Higgs decay channels, we also compute their background cross sections with no Higgs produced. The signal-to-background ratios from the  { cut-based analysis} in the previous section are $\mathcal{O}(10^{-5})$, {$\mathcal{O}(10^{-2})$}, and $\mathcal{O}(10^{-2})$, for the three channels $b\bar{b}$, $W^+W^-$, and $ZZ$, respectively. There is no prominent dependence of these results on the electron beam energy $E_e$. Signal cross sections in $b\bar{b}$ and {$ZZ$} channels are essentially negligible compared with the huge backgrounds. {{$W^+W^-$} channel has an acceptable signal-to-background ratio.  { With the help of the BDT method to improve the cut efficiencies, the signal significances in this channel are brought up to about 1.3$\sigma$, which is still not large enough for identification of the signal in the detector. }  }
 }

{To conclude, it is  { not feasible} to make this complementary precision measurement for $\Gamma(h \to \gamma \gamma)$ at the LHeC using the method presented in this paper, and we} hope that more discriminative methods can be developed to efficiently separate {Higgs photo-production} from  the background processes.

\section{Acknowledgement}
KW would like to thank C.P. Yuan for initial discussion of the project. This work is supported by the National Science Foundation of China (11875232).

\bibliographystyle{utphysmcite}
\bibliography{photo}

\ifx\mcitethebibliography\mciteundefinedmacro
\PackageError{unsrtM.bst}{mciteplus.sty has not been loaded}
{This bibstyle requires the use of the mciteplus package.}\fi
\begin{mcitethebibliography}{10}

\bibitem{Aad:2012tfa}
{ATLAS}, G.~Aad {\em et~al.}, ``{Observation of a new particle in the search
  for the Standard Model Higgs boson with the ATLAS detector at the LHC},''
  \href{http://dx.doi.org/10.1016/j.physletb.2012.08.020}{{\em Phys. Lett.}
  {\bfseries B716} (2012) 1--29},
\href{http://arxiv.org/abs/1207.7214}{{\ttfamily arXiv:1207.7214 [hep-ex]}}.
\mciteBstWouldAddEndPunctfalse
\mciteSetBstMidEndSepPunct{\mcitedefaultmidpunct}
{}{\mcitedefaultseppunct}\relax
\EndOfBibitem
\bibitem{Chatrchyan:2012xdj}
{CMS}, S.~Chatrchyan {\em et~al.}, ``{Observation of a new boson at a mass of
  125 GeV with the CMS experiment at the LHC},''
  \href{http://dx.doi.org/10.1016/j.physletb.2012.08.021}{{\em Phys. Lett.}
  {\bfseries B716} (2012) 30--61},
\href{http://arxiv.org/abs/1207.7235}{{\ttfamily arXiv:1207.7235 [hep-ex]}}.
\mciteBstWouldAddEndPunctfalse
\mciteSetBstMidEndSepPunct{\mcitedefaultmidpunct}
{}{\mcitedefaultseppunct}\relax
\EndOfBibitem
\bibitem{AbelleiraFernandez:2012cc}
{LHeC Study Group}, J.~L. Abelleira~Fernandez {\em et~al.}, ``{A Large Hadron
  Electron Collider at CERN: Report on the Physics and Design Concepts for
  Machine and Detector},''
  \href{http://dx.doi.org/10.1088/0954-3899/39/7/075001}{{\em J. Phys.}
  {\bfseries G39} (2012) 075001},
\href{http://arxiv.org/abs/1206.2913}{{\ttfamily arXiv:1206.2913
  [physics.acc-ph]}}.
\mciteBstWouldAddEndPunctfalse
\mciteSetBstMidEndSepPunct{\mcitedefaultmidpunct}
{}{\mcitedefaultseppunct}\relax
\EndOfBibitem
\bibitem{Han:2009pe}
T.~Han and B.~Mellado, ``{Higgs Boson Searches and the H b anti-b Coupling at
  the LHeC},'' \href{http://dx.doi.org/10.1103/PhysRevD.82.016009}{{\em Phys.
  Rev.} {\bfseries D82} (2010) 016009},
\href{http://arxiv.org/abs/0909.2460}{{\ttfamily arXiv:0909.2460 [hep-ph]}}.
\mciteBstWouldAddEndPunctfalse
\mciteSetBstMidEndSepPunct{\mcitedefaultmidpunct}
{}{\mcitedefaultseppunct}\relax
\EndOfBibitem
\bibitem{Hankele:2006ma}
V.~Hankele, G.~Klamke, D.~Zeppenfeld, and T.~Figy, ``{Anomalous Higgs boson
  couplings in vector boson fusion at the CERN LHC},''
  \href{http://dx.doi.org/10.1103/PhysRevD.74.095001}{{\em Phys. Rev.}
  {\bfseries D74} (2006) 095001},
\href{http://arxiv.org/abs/hep-ph/0609075}{{\ttfamily arXiv:hep-ph/0609075
  [hep-ph]}}.
\mciteBstWouldAddEndPunctfalse
\mciteSetBstMidEndSepPunct{\mcitedefaultmidpunct}
{}{\mcitedefaultseppunct}\relax
\EndOfBibitem
\bibitem{Biswal:2012mp}
S.~S. Biswal, R.~M. Godbole, B.~Mellado, and S.~Raychaudhuri, ``{Azimuthal
  Angle Probe of Anomalous $HWW$ Couplings at a High Energy $ep$ Collider},''
  \href{http://dx.doi.org/10.1103/PhysRevLett.109.261801}{{\em Phys. Rev.
  Lett.} {\bfseries 109} (2012) 261801},
\href{http://arxiv.org/abs/1203.6285}{{\ttfamily arXiv:1203.6285 [hep-ph]}}.
\mciteBstWouldAddEndPunctfalse
\mciteSetBstMidEndSepPunct{\mcitedefaultmidpunct}
{}{\mcitedefaultseppunct}\relax
\EndOfBibitem
\bibitem{Ball:2013hta}
{NNPDF}, R.~D. Ball, V.~Bertone, S.~Carrazza, L.~Del~Debbio, S.~Forte,
  A.~Guffanti, N.~P. Hartland, and J.~Rojo, ``{Parton distributions with QED
  corrections},'' \href{http://dx.doi.org/10.1016/j.nuclphysb.2013.10.010}{{\em
  Nucl. Phys.} {\bfseries B877} (2013) 290--320},
\href{http://arxiv.org/abs/1308.0598}{{\ttfamily arXiv:1308.0598 [hep-ph]}}.
\mciteBstWouldAddEndPunctfalse
\mciteSetBstMidEndSepPunct{\mcitedefaultmidpunct}
{}{\mcitedefaultseppunct}\relax
\EndOfBibitem
\bibitem{Schmidt:2015zda}
C.~Schmidt, J.~Pumplin, D.~Stump, and C.~P. Yuan,
  \href{http://dx.doi.org/10.1103/PhysRevD.93.114015}{``{CT14QED parton
  distribution functions from isolated photon production in deep inelastic
  scattering},''{\em Phys. Rev.} {\bfseries D93} 11, (2016) 114015},
\href{http://arxiv.org/abs/1509.02905}{{\ttfamily arXiv:1509.02905 [hep-ph]}}.
\mciteBstWouldAddEndPunctfalse
\mciteSetBstMidEndSepPunct{\mcitedefaultmidpunct}
{}{\mcitedefaultseppunct}\relax
\EndOfBibitem
\bibitem{Manohar:2016nzj}
A.~Manohar, P.~Nason, G.~P. Salam, and G.~Zanderighi,
  \href{http://dx.doi.org/10.1103/PhysRevLett.117.242002}{``{How bright is the
  proton? A precise determination of the photon parton distribution
  function},''{\em Phys. Rev. Lett.} {\bfseries 117} 24, (2016) 242002},
\href{http://arxiv.org/abs/1607.04266}{{\ttfamily arXiv:1607.04266 [hep-ph]}}.
\mciteBstWouldAddEndPunctfalse
\mciteSetBstMidEndSepPunct{\mcitedefaultmidpunct}
{}{\mcitedefaultseppunct}\relax
\EndOfBibitem
\bibitem{Manohar:2017eqh}
A.~V. Manohar, P.~Nason, G.~P. Salam, and G.~Zanderighi, ``{The Photon Content
  of the Proton},'' \href{http://dx.doi.org/10.1007/JHEP12(2017)046}{{\em JHEP}
  {\bfseries 12} (2017) 046},
\href{http://arxiv.org/abs/1708.01256}{{\ttfamily arXiv:1708.01256 [hep-ph]}}.
\mciteBstWouldAddEndPunctfalse
\mciteSetBstMidEndSepPunct{\mcitedefaultmidpunct}
{}{\mcitedefaultseppunct}\relax
\EndOfBibitem
\bibitem{Chatrchyan:2011ci}
{CMS}, S.~Chatrchyan {\em et~al.}, ``{Exclusive photon-photon production of
  muon pairs in proton-proton collisions at $\sqrt{s}=7$ TeV},''
  \href{http://dx.doi.org/10.1007/JHEP01(2012)052}{{\em JHEP} {\bfseries 01}
  (2012) 052},
\href{http://arxiv.org/abs/1111.5536}{{\ttfamily arXiv:1111.5536 [hep-ex]}}.
\mciteBstWouldAddEndPunctfalse
\mciteSetBstMidEndSepPunct{\mcitedefaultmidpunct}
{}{\mcitedefaultseppunct}\relax
\EndOfBibitem
\bibitem{Budnev:1974de}
V.~M. Budnev, I.~F. Ginzburg, G.~V. Meledin, and V.~G. Serbo, ``{The Two photon
  particle production mechanism. Physical problems. Applications. Equivalent
  photon approximation},''
\href{http://dx.doi.org/10.1016/0370-1573(75)90009-5}{{\em Phys. Rept.}
  {\bfseries 15} (1975) 181--281}.
\mciteBstWouldAddEndPunctfalse
\mciteSetBstMidEndSepPunct{\mcitedefaultmidpunct}
{}{\mcitedefaultseppunct}\relax
\EndOfBibitem
\bibitem{Buckley:2014ana}
A.~Buckley, J.~Ferrando, S.~Lloyd, K.~Nordström, B.~Page, M.~Rüfenacht,
  M.~Schönherr, and G.~Watt, ``{LHAPDF6: parton density access in the LHC
  precision era},''
  \href{http://dx.doi.org/10.1140/epjc/s10052-015-3318-8}{{\em Eur. Phys. J.}
  {\bfseries C75} (2015) 132},
\href{http://arxiv.org/abs/1412.7420}{{\ttfamily arXiv:1412.7420 [hep-ph]}}.
\mciteBstWouldAddEndPunctfalse
\mciteSetBstMidEndSepPunct{\mcitedefaultmidpunct}
{}{\mcitedefaultseppunct}\relax
\EndOfBibitem
\bibitem{Photo}
R.~Li, X.~Lv, B.-W. Wang, and K.~Wang. In preparation.\relax
\mciteBstWouldAddEndPunctfalse
\mciteSetBstMidEndSepPunct{\mcitedefaultmidpunct}
{}{\mcitedefaultseppunct}\relax
\EndOfBibitem
\bibitem{Hirschi:2011pa}
V.~Hirschi, R.~Frederix, S.~Frixione, M.~V. Garzelli, F.~Maltoni, and
  R.~Pittau, ``{Automation of one-loop QCD corrections},''
  \href{http://dx.doi.org/10.1007/JHEP05(2011)044}{{\em JHEP} {\bfseries 05}
  (2011) 044},
\href{http://arxiv.org/abs/1103.0621}{{\ttfamily arXiv:1103.0621 [hep-ph]}}.
\mciteBstWouldAddEndPunctfalse
\mciteSetBstMidEndSepPunct{\mcitedefaultmidpunct}
{}{\mcitedefaultseppunct}\relax
\EndOfBibitem
\bibitem{Alwall:2014hca}
J.~Alwall, R.~Frederix, S.~Frixione, V.~Hirschi, F.~Maltoni, O.~Mattelaer,
  H.~S. Shao, T.~Stelzer, P.~Torrielli, and M.~Zaro, ``{The automated
  computation of tree-level and next-to-leading order differential cross
  sections, and their matching to parton shower simulations},''
  \href{http://dx.doi.org/10.1007/JHEP07(2014)079}{{\em JHEP} {\bfseries 07}
  (2014) 079},
\href{http://arxiv.org/abs/1405.0301}{{\ttfamily arXiv:1405.0301 [hep-ph]}}.
\mciteBstWouldAddEndPunctfalse
\mciteSetBstMidEndSepPunct{\mcitedefaultmidpunct}
{}{\mcitedefaultseppunct}\relax
\EndOfBibitem
\bibitem{Hahn:1998yk}
T.~Hahn and M.~Perez-Victoria, ``{Automatized one loop calculations in
  four-dimensions and D-dimensions},''
  \href{http://dx.doi.org/10.1016/S0010-4655(98)00173-8}{{\em Comput. Phys.
  Commun.} {\bfseries 118} (1999) 153--165},
\href{http://arxiv.org/abs/hep-ph/9807565}{{\ttfamily arXiv:hep-ph/9807565
  [hep-ph]}}.
\mciteBstWouldAddEndPunctfalse
\mciteSetBstMidEndSepPunct{\mcitedefaultmidpunct}
{}{\mcitedefaultseppunct}\relax
\EndOfBibitem
\bibitem{Hahn:2004fe}
T.~Hahn, ``{CUBA: A Library for multidimensional numerical integration},''
  \href{http://dx.doi.org/10.1016/j.cpc.2005.01.010}{{\em Comput. Phys.
  Commun.} {\bfseries 168} (2005) 78--95},
\href{http://arxiv.org/abs/hep-ph/0404043}{{\ttfamily arXiv:hep-ph/0404043
  [hep-ph]}}.
\mciteBstWouldAddEndPunctfalse
\mciteSetBstMidEndSepPunct{\mcitedefaultmidpunct}
{}{\mcitedefaultseppunct}\relax
\EndOfBibitem
\bibitem{Sjostrand:2006za}
T.~Sjostrand, S.~Mrenna, and P.~Z. Skands, ``{PYTHIA 6.4 Physics and Manual},''
  \href{http://dx.doi.org/10.1088/1126-6708/2006/05/026}{{\em JHEP} {\bfseries
  05} (2006) 026},
\href{http://arxiv.org/abs/hep-ph/0603175}{{\ttfamily arXiv:hep-ph/0603175
  [hep-ph]}}.
\mciteBstWouldAddEndPunctfalse
\mciteSetBstMidEndSepPunct{\mcitedefaultmidpunct}
{}{\mcitedefaultseppunct}\relax
\EndOfBibitem
\bibitem{deFavereau:2013fsa}
{DELPHES 3}, J.~de~Favereau, C.~Delaere, P.~Demin, A.~Giammanco, V.~Lemaître,
  A.~Mertens, and M.~Selvaggi, ``{DELPHES 3, A modular framework for fast
  simulation of a generic collider experiment},''
  \href{http://dx.doi.org/10.1007/JHEP02(2014)057}{{\em JHEP} {\bfseries 02}
  (2014) 057},
\href{http://arxiv.org/abs/1307.6346}{{\ttfamily arXiv:1307.6346 [hep-ex]}}.
\mciteBstWouldAddEndPunctfalse
\mciteSetBstMidEndSepPunct{\mcitedefaultmidpunct}
{}{\mcitedefaultseppunct}\relax
\EndOfBibitem
\bibitem{Lapertosa:2016zpo}
A.~Lapertosa, ``{Calibration of the b-tagging efficiency on jets with charm
  quark for the ATLAS experiment},'' Master's thesis, Genoa U.\relax
\mciteBstWouldAddEndPunctfalse
\mciteSetBstMidEndSepPunct{\mcitedefaultmidpunct}
{}{\mcitedefaultseppunct}\relax
\EndOfBibitem
\bibitem{Barger:1990mn}
V.~D. Barger, G.~Bhattacharya, T.~Han, and B.~A. Kniehl, ``{Intermediate mass
  Higgs boson at hadron supercolliders},''
\href{http://dx.doi.org/10.1103/PhysRevD.43.779}{{\em Phys. Rev.} {\bfseries
  D43} (1991) 779--788}.
\mciteBstWouldAddEndPunctfalse
\mciteSetBstMidEndSepPunct{\mcitedefaultmidpunct}
{}{\mcitedefaultseppunct}\relax
\EndOfBibitem
\bibitem{Hocker:2007ht}
A.~Hocker {\em et~al.}, ``{TMVA - Toolkit for Multivariate Data Analysis},''
  \href{http://arxiv.org/abs/physics/0703039}{{\ttfamily
  arXiv:physics/0703039}}\relax
\mciteBstWouldAddEndPunctfalse
\mciteSetBstMidEndSepPunct{\mcitedefaultmidpunct}
{}{\mcitedefaultseppunct}\relax
\EndOfBibitem
\end{mcitethebibliography}

\end{document}